\documentstyle[12pt]{article}
\newcounter{eqn}
\def\lab{\refstepcounter{eqn}\eqno(\arabic{eqn})}
\def\l#1{\lab\label{#1}}
\def\r#1{(\ref{#1})}
\begin{document}
\begin{titlepage}
\title{LARGE-N EXPANSION AS SEMICLASSICAL APPROXIMATION
TO THE THIRD-QUANTIZED THEORY}

\author{
 V.P.Maslov and  O.Yu. Shvedov
\\
{\small {\em Chair of Quantum Statistics and Field Theory,}}\\
{\small{\em Department of Physics, Moscow State University }}\\
{\small{\em Vorobievy gory, Moscow 119899, Russia}} }

\end{titlepage}
\maketitle

\begin{flushright}
hep-th/9805089
\end{flushright}

\footnotetext{e-mail: olshv@ms2.inr.ac.ru,
shvedov@qs.phys.msu.su}

\begin{center}
{\bf Abstract}
\end{center}

\begin{flushright}

\parbox{17cm}{
The semiclassical theory for the large-$N$ field models  is  developed
from an  unusual  point  of view.  Analogously to the procedure of the
second quantization in quantum mechanics,  the functional  Schrodinger
large-$N$ equation   is  presented  in  a  third-quantized  form.  The
third-quantized creation and  annihilation  operators  depend  on  the
field $\varphi({\bf  x})$.  If the coefficient of the $\varphi^4$-term
is of order $1/N$ (this is a usual condition of applicability  of  the
$1/N$-expansion), one  can  rescale  the  third-quantized operators in
such a way that their commutator will be small,  while the  Heisenberg
equations will not contain large or small parameters.  This means that
classical equation  of  motion  is  an  equation  on  the   functional
$\Phi[\varphi(\cdot)]$. This  equation being a nonlinear analog of the
functional Schrodinger  equation   for   the   one-field   theory   is
investigated. The    exact   solutions   are   constructed   and   the
renormalization problem is analysed.  We also perform  a  quantization
procedure about   found   classical   solutions.   The   corresponding
semiclassical theory is a theory of a variable number of  fields.  The
developed third-quantized  semiclassical  approach  is  applied to the
problem of finding the large-$N$ spectrum.  The results  are  compared
with obtained  by known methods.  We show that not only known but also
new energy levels can be found.
}

\end{flushright}
 
\newpage

\section{Introduction}

Investigation of quantum field theory as the number of fields tends to
infinity is very important. For example, the number of colours of
quarks $N_c$ seems to be the only possible large parameter in 
quantum chromodynamics, so that one can perform asymptotic expansions
in a series of $1/N_c$ \cite{tHooft1,Witten}.

Vector models are also suitable for the large-N analysis. Expansions in
a series of $1/N_f$ ($N_f$ - number of flavours of quarks) is widely
used in QCD. Examples of applications of the $1/N$-expansion are:

- large-N calculation of Green functions \cite{Wilson};

- evaluation of the effective action \cite{CJT};

- investigations of the spontaneous symmetry breaking 
\cite{CJP,GN};

- research of processes of particle creation and back reaction in the
strong external field \cite{CM1};

- investigation of the evolution of the pion condensate \cite{CM2};

- evaluations of cross-sections of the processes like ``1 particle''
$\to$ ``n particles'' \cite{Mak}.

There are various approaches to the $1/N$-expansion.
Consider, for example, the simplest vector model
$$
{\cal L} = \frac{1}{2} \partial_{\mu}\varphi^a
\partial_{\mu}\varphi^a - \frac{m^2}{2} \varphi^a\varphi^a
-\frac{\lambda}{4N}\varphi^a\varphi^a\varphi^b\varphi^b
\l{l}
$$
($a,b=1,...,N$, $\varphi^1,...,\varphi^N$ are scalar fields, we
sum over repeated indices).
Some of these approaches are the following.

1. One can extract the so-called bubble Feynman graphs
and evaluate their contributions to the Green functions and
scattering amplitudes \cite{Wilson,Witten}; one can then show that
contibution of other graphs can be neglected.

2. One can present physical quantities via  functional  integrals  like
$$
\int D\varphi e^{i\int dx {\cal L}},
\l{int}
$$
introduce the auxiliary field $\chi$ \cite{CJP,CM3,A3}, insert the
identity
$$
1= \int D\chi e^{\frac{i\lambda}{4N}\int dx (N\chi-\varphi^a\varphi^a)^2}
$$
into the functional integral \r{int} and evaluate the Gaussian integral
over $\varphi$ exactly. Then remaining integral over $\chi$ is to be
evaluated semiclassically (by the saddle-point technique).  It is  the
condition $N\to\infty$  that allows us to apply the semiclassical
approach.

3. One can use the collective field approach \cite{JP}. The
Schrodinger wave functional
$\Psi[\varphi_1(\cdot),...,\varphi_N(\cdot)]$ can be assumed to depend
only on the collective field
$$
\psi({\bf x},{\bf y})=\sum_{a=1}^N \varphi^a({\bf x}) \varphi^a({\bf
y}).
\l{psi}
$$
The ansatz
$$
\Psi[\varphi_1(\cdot),...,\varphi_N(\cdot)]=\Psi[\psi(\cdot,\cdot)]
\l{psi2}
$$
satisfies the Schrodinger equation; the obtained relation for
$\Psi[\psi(\cdot,\cdot)]$ can br treated semiclassically.

4. One can develop the following semiclassical approach
\cite{BY}. Consider the operators
$$
\hat{\psi}({\bf x},{\bf y})= \sum_{a=1}^N \hat{\varphi}^a({\bf x})
\hat{\varphi}^a({\bf y}),
$$
$$
\hat{\xi}({\bf x},{\bf y})= \sum_{a=1}^N \hat{\pi}^a({\bf x})
\hat{\varphi}^a({\bf y}),
$$
$$
\hat{\eta}({\bf x},{\bf y})= \sum_{a=1}^N \hat{\pi}^a({\bf x})
\hat{\pi}^a({\bf y}).
$$
The system of Heisenberg equations for these operators is expressed
via these operators only, so that one can use the argumentation like
the Ehrenfest theorem in quantum mechanics. One can consider the average
values of the right- and left-hand sides of the Heisenberg equations
and obtain the classical equations.

5. One can consider the Heisenberg equations for the fields
$\varphi^a$. Making use of the averaging procedure \cite{CM4}, one can
obtain the large-N equations.

In this  paper we present yet another approach to the $1/N$-expansion.
We show that semiclassical investigation of the large-N  theories  may
lead to  classical  equations of another type rather than to equations
on  ``collective field'' $\psi({\bf x},{\bf y})$ \r{psi} or
$\varphi^a$.

We consider states of a more general form than $\Psi[\psi(\cdot,\cdot)]$.
The state functional
$$
\Psi[\varphi_1(\cdot),...,\varphi_N(\cdot)]
\l{st}
$$
is considered to be  symmetric  with  respect  to  transformations  of
fields $\varphi_1({\bf x})$,  ...,  $\varphi_N({\bf x})$.  We will also
discuss the general, non-symmetric, case.

In quantum statistical mechanics the most suitable way to  investigate
the $N$-particle   Schrodinger   equation  is  the  method  of  second
quantization. The Hamiltonian operator is presented via the  operators
of creating  and  annihilating  particles,  $a^{\pm}(q)$,  with  given
coordinate $q$. Making use of the canonical commutation relations, one
can investigate different properties of the $N$-particle system.

One can hope that this conception can be applied also to the $N$-field
system. The $N$-field Hamiltonian operator is to be presented via  the
operators $A^{\pm}[\varphi(\cdot)]$ which ``create'' or ``annihilate''
the field configuration $\varphi(\cdot)$. This procedure can be called
as ``third   quantization''   because   quantum   field   is   already
second-quantized.

The notion of third quantization usually arise in quantum gravity  and
cosmology \cite{3q} when the processes (like wormhole
transition) with variable number of universes  are  studied.  Creation
(annihilation) operators create (annihilate) the universe.

The conception  of third quantization is also useful in string theory:
splitting and joining of strings are interpretted in terms of creation
and annihilation operators \cite{KK}.

We see  that  the  idea  of  third quantization can be applied also to
ordinary field theories.  We will show that this idea is  very  useful
and allows us to construct new asymptotics in large-N field theories.

For example, one can investigate the processes with a large number $n$
of particles, $n \sim N$, by using the developed technique. It is known
\cite{Mak} that  the  usual  $1/N$-expansion  is  applicable  only for
processes with a small number of particles.

We will  consider  the formalism of third quantization in more details
in section 2. We will show that the coefficient of $\varphi^4$ in
eq.\r{l} is  of  order  $O(1/N)$  is  a  condition of applicability of
semiclassical approximation  to  the   third-quantized   theory.   The
corresponding classical equations are unusual.  Namely,  the classical
equations corresponding to quantum mechanics are ordinary differential
equations for the classical trajectory $(p(t),q(t))$, $p(t)$ are momenta,
$q(t)$ are coordinates.  When one applies semiclassical approximation to
the (second-quantized) field theory, one finds classical equations for
the classical field $\varphi({\bf  x},  t)$.  In  the  third-quantized
theory the   classical   equation   is  equation  for  the  functional
$\Phi[t,\varphi(\cdot)]$ depending on the field $\varphi({\bf x})$ and
time $t$.  Although  this  equation  resembles  the  ordinary  quantum
functional Schrodinger  equation  in  the  theory  of  one  field,  an
essential feature   of   this   equation   is  non-linearity.  Another
interesting feature of the equation on $\Phi[t,\varphi(\cdot)]$ is
exact solvability for some interesting cases.

The conception   of  classical  ``master  field''  is  often  used  in
investigation of large-N systems. However, ``master field'' is usually
considered as  a function on a finite-dimensional space.  For example,
field $\psi({\bf x},{\bf y})$ \r{psi} or fields  $\varphi^a({\bf  x})$
are ususally  considered  as  classical variables \cite{BY,
CM4,CM5,MF}. We see that in our approach the role of ``master
field'' is played by the functional $\Phi[t,\varphi(\cdot)]$ which can
be treated as time-dependent vector of the one-field  Fock  space.
Classical mechanics for $\psi({\bf x},{\bf y})$ is in
agreement with our dynamics for $\Phi$.  Namely,  the Gaussian ansatz
for our  equation for $\Phi$  leads to the equation on $\psi$ which
was obtained in refs.\cite{JP}.  However,  our classical theory  is
much more  rich than the mechanics for $\psi$ because one can consider
other solutions for the equation for $\Phi$. The reason is that we are
considering
solutions to the $N$-field equation
which are   not   $O(N)$-symmetric   but  symmetric  with  respect  to
transpositions.  This class
is much more wide than the class of  solutions  \r{psi2}.  This  means
that we  are able not only to reproduce the known asymptotics but also
to construct new asymptotic solutions  to  the  large-N  quantum-field
Schrodinger equation.

An interesting    feature    of    the    classical    equations    on
$\Phi[\varphi(\cdot)]$ is that it can be interpretted as a Hamiltonian
system corresponding to the flat phase space.  On the other hand,  the
phase space of refs.\cite{BY} is curved. There are a lot of
well-developed semialassical  methods  \cite{M1,M2,sol,
M3,Gt}  for
flat phase  space,  so  that  all  of  them  are  applicable  for  the
third-quantized system.

We will   construct  the  asymptotics  for  the  entire  N-field  wave
functional \r{st}.  This approach  gives  more  information  than  the
averaging procedure of the Heisenberg field equations.  The difference
can be explained as follows.  Consider the semiclassical approximation
to the ordinary quantum mechanics. One can derive classical equations with
the help of the Ehrenfest theorem.  This result allows us to find  the
trajectory of propagation of the wave packet. To find the evolution of
the shape  of  the  wave  packet,  it  is  necessary  to  apply   more
complicated semiclassical  methods  such  as  the  complex-WKB  method
\cite{M2} which allows us to construct the approximation for the  wave
function rather   than   for  average  values  of  the  semiaclassical
observables.

This paper is organized as follows.  Section 2 deals with  the  simple
quantum mechanical model,  the $O(N)$-symmetric anharmonic oscillator.
One can test different versions of the $1/N$-expansion on such  a  toy
model. We consider the analogs of the collective field approach and of
our approach.  These methods are compared,  and it is shown  that  our
approach allows  us  to  construct  much  more rich set of approximate
solutions to the Schrodinger equation.  In section 3 we introduce  the
notion of third quantization (as related to the large-$N$ system).  We
write down the third-quantized quantum Hamiltonian,  as  well  as  the
Heisenberg equations  of  motion.  The  semiclassical approximation is
applied and the problems of  regularization  and  renormalization  are
considered. In  section 4 we apply the developed method to the problem
of investigation of the large-$N$ field theory in the  finite  volume.
Section 5  contains  concluding remarks.  Appendices A and B deal with
the brief review of the complex-WKB technique \cite{M2,MS1}.

\section{$O(N)$ - symmetric anharmonic oscillator: different
approaches to $1/N$-expansion}

In this section we consider the simplest example of the large-N
system - the $O(N)$-symmetric anharmonic oscilator with quartic
interaction. The wave function of this system depends on $N$
coordinates, $x_1$,...,$x_N$, while the Hamiltonian is
$$
H=\sum_{i=1}^N \left(-\frac{1}{2}
\frac{\partial^2}{\partial x_i^2}+\frac{ax_i^2}{2}
\right)+ \frac{g}{4N}\sum_{ij=1}^N x_i^2x_j^2.
\l{anh}
$$
Different approaches to $1/N$-expansion can be illustrated by applying
to such a simple system. Let us compare the third-quantized approach
developed in this paper with other approaches.

\subsection{The collective field approach}

The collective field approach for the Hamiltonian \r{anh} which is
analogous to the ansatz \r{psi2} was developed in \cite{JP}. The idea
is to consider the $O(N)$-symmetric wave function depending only on
$r=\sqrt{x_1^2+...+x_N^2}$,
$$
\psi_N(x_1,...,x_N)=\psi(\sqrt{x_1^2+...+x_N^2}).
$$
However, the full probability will not have the usual form $\int dr
|\psi(r)|^2$, since the area of the surface of the sphere depends on its
radius. This means that
$$
\int dx_1...dx_N |\psi_N(x_1,...,x_N)|^2= C_N \int_0^{\infty} dr
|\psi(r)|^2 r^{N-1},
$$
where $C_N$ is an $r$-independent constant. Let us consider the
quantity
$$
\varphi(r)=\psi(r) r^{\frac{N-1}{2}},
\l{phi}
$$
which plays the role of the probability amplitude that
$r=\sqrt{x_1^2+...+x_N^2}$. Time evolution of the function \r{phi}
$\varphi^t(r)$ is specified by the equation derivable from
eq.\r{anh}:
$$
i\frac{\partial \varphi}{\partial t} =
-\frac{1}{2}\frac{\partial^2\varphi}{\partial r^2}+
\frac{N^2-1}{8r^2}\varphi + \frac{ar^2}{2}\varphi +
\frac{g}{4N}r^4 \varphi
\l{a2w}
$$
The rescaling
$$
r=\xi\sqrt{N}
$$
transforms eq.\r{a2w} to the semiclassical form
$$
\frac{i}{N}\frac{\partial \varphi}{\partial t}=
-\frac{1}{2N^2}\frac{\partial^2\varphi}{\partial\xi^2}+
\frac{1}{8\xi^2} \left(1-\frac{1}{N^2}\right) \varphi +
\frac{a\xi^2}{2}\varphi + \frac{g\xi^4}{4} \varphi,
\l{a3}
$$
where the analog of the Planck constant is $1/N$. The condition
$N\to\infty$ is the condition of applicability of semiclassical
methods to eq.\r{a3}.

Eq.\r{a3} specifies the motion of the particle in the potential
$$
U=\frac{1}{8\xi^2}+\frac{a\xi^2}{2}+\frac{g\xi^4}{4}.
\l{u1}
$$
 One can apply different semiclassical methods to
eq.\r{a3}
which are described in Appendix A.
Spectrum in the vicinity
of the ground state can be found by using the oscillator
approximation of refs.\cite{sol, M2,M4}:
$$
\frac{E_n}{N}=U(\overline{\xi})+ \frac{1}{N}\sqrt{U''(\overline{\xi})}
(n+1/2) + O(1/N^2),
\l{x1}
$$
where $\overline{\xi}$ is the minimum of the potential \r{u1}.

\subsection{The ``second-quantized'' approach}

Let us develop for the $O(N)$ - symmetric system the approach based on
the Fock space. Instead of the Schrodinger equation for
$\psi_N(x_1,...,x_N)$, we consider the evolution equation for the set
of functions
$$
(\psi_0,\psi_1(x_1),...,\psi_N(x_1,...,x_N),...)
\l{f2w}
$$
($\psi_N$ being symmetric). Introduce in the Fock space of sets \r{f2w}
creation and annihilation operators as usual,
$$
(a^-(x)\psi)_{n-1}(x_1,...,x_{n-1})=\sqrt{n}\psi_n(x_1,...,x_n),
$$
$$
(a^+(x)\psi)_n(x_1,...,x_n)= \frac{1}{\sqrt{n}}\sum_{i=1}^n
\delta(x-x_i) \psi_{n-1} (x_1,...,x_{i-1},x_{i+1},...,x_n).
$$
Since the operator
$$
\sum_{i=1}^n x_i^2
$$
is presented via the introduced operators as
$$
\int dx a^+(x) x^2 a^-(x),
$$
while for the operator $\sum_{i=1}^n (-\partial^2/\partial x_i^2) $
one has
$$
\int dx a^+(x) \left(-\frac{\partial^2}{\partial x^2}\right) a^-(x),
$$
one can present the Schrodinger equation in the second-quantized form
$$
i\frac{\partial \psi}{\partial t} = \hat{H}\psi
\l{s3}
$$
where
$$
\hat{H}=\int dx a^+(x) \left(-\frac{1}{2}\frac{\partial^2}{\partial x^2}
+\frac{ax^2}{2}\right)a^-(x)+ \frac{g\varepsilon}{4}
\left(\int dx a^+(x)x^2a^-(x)\right)^2,
\l{h3}
$$
for $\varepsilon=1/N$. Since the operators $a^{\pm}(x)$ obey canonical
commutation relations
$$
[a^{\pm}(x),a^{\pm}(y)]=0, \qquad [a^-(x),a^+(y)]=\delta(x-y),
$$
one can use the following functional Schrodinger representation. The
vacuum state of the Fock space (1,0,0,...) corresponds to the
functional
$$
\Psi(\xi)=\exp\left(-\frac{1}{2\varepsilon} \int dx \xi^2(x)\right)
$$
while creation and annihilation operators are written as
$$
a^{\pm}(x)=\frac{\xi(x) \mp \varepsilon \delta/\delta \xi(x)}
{\sqrt{2\varepsilon}}.
$$
In this representation eq.
\r{s3} takes the semiclassical form
$$
i\varepsilon \frac{\partial \psi[\xi]}{\partial t}
= {\cal H} \left[\xi, -i\varepsilon \frac{\delta}{\delta \xi}\right]
\Psi[\xi(\cdot)],
\l{s4}
$$
where
$$
{\cal H}[\xi,\pi]=
H\left[\frac{\xi-i\pi}{\sqrt{2}},\frac{\xi+i\pi}{\sqrt{2}}\right],
\l{x2}
$$
$$
H[\varphi^*,\varphi]=\int dx \varphi^*(x) \left(-\frac{1}{2}
\frac{\partial^2}{\partial x^2}+ \frac{ax^2}{2}\right)\varphi(x)+
\frac{g}{4} \left(\int dx x^2|\varphi(x)|^2\right)^2.
$$
An analog of the Planck constant is $\varepsilon$.
One can try to apply semiclassical methods to eq.\r{s4}. One finds
that the corresponding classical equations
$$
\frac{d}{dt}\xi=\frac{\delta {\cal H}}{\delta \pi},
\frac{d}{dt}\pi=-\frac{\delta {\cal H}}{\delta \xi}
\l{x3}
$$
can be presented as
$$
i\frac{d}{dt}\varphi=
\frac{\delta H}{\delta \varphi^*}, \qquad \varphi=
\frac{\xi+i\pi}{\sqrt{2}}.
\l{x4}
$$
We see that classical trajectory in our approach is specified not by
one coordinate and one momentum but by infinite set of coordinates and
momenta, $(\xi(\cdot),\pi(\cdot))$, or, equivalently, by the complex
function $\varphi(t,x)$ obeying the following dynamical equation
$$
i\frac{\partial \varphi(t,x)}{\partial t} = (-\frac{1}{2}
\frac{\partial^2}{\partial x^2}+ \frac{ax^2}{2})\varphi(t,x)+
\frac{g}{2}x^2\varphi(t,x) \int dy y^2 |\varphi(t,y)|^2.
\l{rt1}
$$
The reason for extending the phase space is the following. First, we
consider not the ``N-argument'' Schrodinger equation but infinite sets
of such equations for $N=\overline{0,\infty}$. The average value of
``number of arguments of wave function'' analogously to the average
number of particles in quantum statistics
$$
<\int dx a^+(x)a^-(x)>= \frac{1}{2\varepsilon}
<\int dx \left(\xi(x)-\varepsilon \frac{\delta}{\delta \xi(x)}\right)
\left(\xi(x)+\varepsilon \frac{\delta}{\delta \xi(x)}\right)>
$$
in the ``semiclassical state'' is approximately equal to
$$
\frac{1}{2\varepsilon} \int dx (\xi(x)-i\pi(x))
(\xi(x)+i\pi(x))= \frac{1}{\varepsilon}\int dx|\varphi(x)|^2.
$$
Because of the relation $\varepsilon=1/N$ one should impose on the
function $\varphi$ the normalization condition
$$
\int dx |\varphi(x)|^2 = 1.
\l{n1}
$$

However, the extension of the state space to the Fock space \r{f2w} is
not the only reason for extension of the phase space. Another reason
is that we consider not only $O(N)$ - symmetric solutions to the
Schrodinger equation for $\psi_N$ but solutions of a more general form.

\subsection{Correspondence between Hamiltonian systems}

Let us show that the collective-field equations can be
obtained as a partial case of our classical equation \r{rt1}.
To investigate this problem, consider the substitution
$$
\varphi(t,x)=c e^{\frac{i}{2}\alpha x^2}
\l{a1w}
$$
to eq.\r{rt1} for $t$-dependent complex numbers $c$ and $\alpha$.
Making use of the condition \r{n1}, we find that the function \r{a1w}
obeys eq.\r{rt1} if
$$
i\frac{d}{dt}c = -\frac{i\alpha}{2},
\l{hs1w}
$$
$$
\frac{d\alpha}{dt}+\alpha^2 + a + \frac{g}{2Im \alpha}=0
\l{hs2w}
$$
However, eq.\r{hs2w} can be transformed to the form of the Hamiltonian
system corresponding to the potential \r{u1}. Let us extract real and
imaginary parts of $\alpha$,
$$
p=\frac{Re \alpha}{\sqrt{2 Im \alpha}},
\qquad
x=\frac{1}{\sqrt{2 Im \alpha}},
$$
Eq.\r{hs2w} takes the form
$$
\frac{dx}{dt}=p, \qquad \frac{dp}{dt}=-\frac{\partial U}{\partial x}.
$$
We see that our approach contains the collective field approach as a
partial case. However, one can consider not only the Gaussian ansatz
\r{a1w} for eq.\r{rt1} but also other substitutions. We see that one
can construct new asymptotic solutions for the large-N systems.

\subsection{Asymptotic spectrum of the second-quantized Hamiltonian}

Let us  show  that  the  oscillator  approximation  applied   to   the
Hamiltonoan system with the Hamiltonian \r{x2} leads us to the
semiclassical spectrum of energy which is much more rich than
eq.\r{x1}.

There are many semiclassical methods. One of them - quantization of
periodic trajectories - is the following
\cite{sol}. One should find the
classical periodic solution. Then one considers small variations
around it and finds the ``stability angles'' \cite{sol} related with
the distances between semiclassical energy levels. This approach is
certainly applicable to the Hamiltonian system \r{x2} since there are
many periodic solutions to eq.\r{rt1} of the form
$$
\varphi(t,x)=\varphi(x)e^{-i\Omega t}.
\l{x4a}
$$
However, we will apply the much more simple approach - quantization of
static solutions. To apply this approach, it is necessary \cite{KcM}
to consider the Hamiltonian
$$
\hat{H}_{\Omega}=\hat{H}-\Omega \int dx a^+(x)a^-(x)
\l{x41}
$$
instead of the operator \r{h3}. Since the spectra of the $N$-particle
parts of these Hamiltonians are related by shifting by $\Omega N$, one
can investigate the Hamiltonian \r{x41} instead of \r{h3}.

The classical Hamiltonian function corresponding to the operator
\r{x41} has the form \r{x2}, where
$$
H_{\Omega}[\varphi^*,\varphi]= \int dx \varphi^*(x)
\left(-\frac{1}{2}\frac{\partial^2}{\partial x^2} + \frac{ax^2}{2} -
\Omega\right)\varphi(x) +       \frac{g}{4}       \left(\int        dx
x^2|\varphi(x)|^2\right)^2.
\l{x42}
$$
Let us find the .static solutions of the classical equation of motion
\r{x4} which takes the form of the stationary harminic oscillator
equation
$$
\left(-\frac{1}{2}\frac{\partial^2}{\partial x^2} +
\frac{\omega^2x^2}{2} -
\Omega\right)\varphi(x)=0,
\l{x43}
$$
where the frequency $\omega$ is expressed via the average value of
$x^2$:
$$
\omega^2 = a+ g\int dx x^2 |\varphi(x)|^2.
\l{x44}
$$
Note that the same equation \r{x43} could be obtained by substitution
of periodic solution \r{x4a} to the equation \r{rt1}.

The solution to eq.\r{x43} can be chosen as the $K$-th excited state
of the harmonic oscillator:
$$
\varphi = \Psi_K = \frac{(A^+)^K}{\sqrt{K!}}\Psi_0,
$$
where
$$
\Psi_0= const e^{-\frac{\omega x^2}{2}}
$$
is the ground oscillator state, while
$$
A^+ = \sqrt{\frac{\omega}{2}}x - \frac{1}{\sqrt{2\omega}}
\frac{\partial}{\partial x}
$$
is the creation operator. The parameter $\Omega$ is
$$
\Omega = \omega (K+1/2).
$$
The normalizing factor is determined by eq.\r{n1}.
Since the operator of coordinate can be presented via creation and
annohilation operators,
$$
x=\frac{1}{\sqrt{2\omega}}(A^++A^-)
\l{x4b}
$$
one can find the average value
$$
(\Psi_K, x^2 \Psi_K) = \frac{1}{\omega}(K+1/2),
$$
so that the equation on $\omega$ \r{x42} can be presented as
$$
\omega^2 = a+ \frac{g}{\omega}
(K+1/2).
$$

Let us quantize the constructed static solution. It is necessary to
consider the small perturbations around it:
$$
\xi(x) \to \xi(x)+\delta \xi(x,t),
\qquad\pi(x) \to \pi(x)+\delta \pi(x,t).
$$
Then one should consider such variations that
$$
\delta\xi (x,t) = \delta\xi(x) e^{i\beta t},\qquad
\delta\pi (x,t) = \delta \pi(x) e^{i\beta t}.
$$
This means that variations of classical variables $\xi$ and $\pi$
should be complex, although the classical variables are real. It is
convenient to make a transformation:
$$
\delta\varphi = \frac{\delta\xi+i\delta\pi}{\sqrt{2}},\qquad
\delta\varphi^* = \frac{\delta\xi-i\delta\pi}{\sqrt{2}}
$$
The fact of complexity $\delta\xi$ and $\delta\pi$ means that the
variations of $\varphi$ and $\varphi^*$ are not conjugated each other;
they are independent. Thus, to investigate the small variations around
static solution, one should consider the Hamiltonian system
$$
i\frac{\partial \varphi}{\partial t} =
\frac{\delta H_{\Omega}}{\delta \varphi^*},\qquad
-i\frac{\partial \varphi^*}{\partial t} =
\frac{\delta H_{\Omega}}{\delta \varphi},
$$
consider the independent perturbations of $\varphi$ and $\varphi^*$:
$$
\varphi \to \varphi + Fe^{i\beta t},\qquad
\varphi^* \to \varphi^* + Ge^{i\beta t},
$$
and find a spectrum of $\beta$:
$$
-\beta F = \frac{\delta^2H_{\Omega}}{\delta\varphi^*\delta\varphi}F+
\frac{\delta^2H_{\Omega}}{\delta\varphi^*\delta\varphi^*}G,
$$
$$
\l{x45}
$$
$$
\beta G = \frac{\delta^2H_{\Omega}}{\delta\varphi\delta\varphi}F+
\frac{\delta^2H_{\Omega}}{\delta\varphi\delta\varphi^*}G,
$$
which should be real if the static solution is stable. The more
detailed derivation of the semiclassical spectrum is presented in
Appendix B.

The variation system \r{x45} has the following form for the
Hamiltonian \r{x42}:
$$
-\beta F(x)=
\left(-\frac{1}{2}\frac{\partial^2}{\partial x^2}
+ \frac{\omega^2x^2}{2} -
\Omega\right)F(x) + \frac{g}{2}x^2\varphi^2(x) \int dy y^2 \varphi(y)
(F+G)(y),
$$
$$
\l{x46}
$$
$$
\beta G(x)=
\left(-\frac{1}{2}\frac{\partial^2}{\partial x^2}
+ \frac{\omega^2x^2}{2} -
\Omega\right)G(x)  + \frac{g}{2}x^2\varphi^2(x) \int dy y^2 \varphi(y)
(F+G)(y),
$$
The solutions to eq.\r{x46} are presented as follows. First of all,
let us try to substitute instead of $F$ and $G$ the $n$-th
eigenfunction of the oscillator:
$$
F^+_n=0, G^+_n=\Psi_n,
\l{x47}
$$
or
$$
F^-_n=\Psi_n, G^-_n=0.
\l{x48}
$$
Since the matrix element $(\Psi_K,x^2 \Psi_n)=0$ if $n-K\ne 0, \pm 2$,
functions \r{x47}, \r{x48} obeys eq.\r{x46} if
$$
\beta_n^+ = \omega (n-K),
\beta_n^- = \omega (K-n).
$$
Let us find other solutions to eq.\r{x46}. First of all, there is a
``zero-mode'' solution corresponding to the invariance of the
Hamiltonian system with respect to transformations $\varphi \to
\varphi e^{i\alpha}$, $\varphi^* \to \varphi^* e^{-i\alpha}$:
$$
F=-\Psi_K, G=\Psi_K, \beta=0.
$$
Further, making use of eq.\r{x4b}, one has
$$
x^2\varphi= \frac{1}{2\omega}
(\sqrt{(K+1)(K+2)}\Psi_{K+2}+(2K+1)\Psi_K+\sqrt{K(K-1)}\Psi_{K-2})
$$
Thus, we find the following two solutions to system \r{x46}:
$$
F=\sqrt{K(K-1)}\Psi_{K+2}, G=-\sqrt{(K+1)(K+2)}\Psi_{K-2},
\beta=-2\omega;
$$
$$
G=\sqrt{K(K-1)}\Psi_{K+2}, F=-\sqrt{(K+1)(K+2)}\Psi_{K-2},
\beta=2\omega
$$
because for these cases $(x^2\Psi_K,F+G)=0$.

The non-trivial solution to eq.\r{x46} is
$$
F= -\frac{\sqrt{(K+1)(K+2)}}{\beta+2\omega} \Psi_{K+2}
-\frac{\sqrt{K(K-1)}}{\beta-2\omega} \Psi_{K-2},
$$
$$
G= \frac{\sqrt{(K+1)(K+2)}}{\beta-2\omega} \Psi_{K+2}
+\frac{\sqrt{K(K-1)}}{\beta+2\omega} \Psi_{K-2},
$$
where
$$
\beta^2=4\omega^2 + \frac{g}{\omega}(2K+1).
$$
There is also a solution to the non-stationary variation system which
is proportional to $t$:
$$
\delta \varphi^*-\delta\varphi = 2i\omega(2K+1)t \Psi_K,
$$
$$
\delta \varphi^*+\delta\varphi= - \sqrt{(K+1)(K+2)}\Psi_{K+2}
+\sqrt{K(K-1)}\Psi_{K-2} + (2+\frac{8\omega^3}{(2K+1)g})\Psi_K.
$$
However, such linear instability of classical static solution is usual
for quantum systems with zero-modes. There are several approaches to
resolve the difficulty. It is shown in Appendix B that such an
instability is not an obstacle for quantization.

The asymptotic energy spectrum for the quantum Hamiltonian
$H_{\Omega}$ is expressed via frequences $\beta$:
$$
E^n_{\Omega} = N{\cal E}_{\Omega} + \varepsilon_0 +
\sum_m \beta_m n_m +O(1/N),
\l{x4e1}
$$
where $E_{\Omega}^n$ is a quantum energy, ${\cal E}_{\Omega}$ is a
corresponding ``classical'' energy,
$$
{\cal E}_{\Omega}=H_{\Omega}(\varphi^*,\varphi),
$$
$\varepsilon_0$ is a quantum
correction, $\beta_m$ are energy excitations. Non-negative numbers
$n_m$ are parameters specifying the energy levels.

It is important that not all values of $\beta_m$ found above enters to
eq.\r{x4e1}. When we quantize ordinary harmonic oscillator (or free
quantum field), there are positive-frequency and negative-frequency
solutions to the variation equations (for free fields they coincide
with classical equations). However, one usually takes into account only
positive frequences to construct a spectrum.

It happens (see appendix B) that in our case we should take into
account not positive values of $\beta$ but such values of $\beta$ that
obey the following condition:
$$
(G,G)>(F,F).
$$
This means that the excitation energies entering to eq.\r{x4e1} are:
$$
\beta_m=\omega (m-K), m\ne K,K\pm 2,
$$
$$
\beta_{K-2} = -2\omega,
$$
$$
\beta_{K+2} = + \sqrt{4\omega^2 + \frac{g}{\omega}(2K+1)}.
$$
An asymptotic energy spectrum of the operator $\hat{H}$ which
corresponds to the pertiodic solution \r{x4a} has then the form
$E=E_{\Omega}+N\Omega$, or:
$$
E^n = N{\cal E} + \varepsilon_0 +
\sum_{m\ge 0,m\ne K} \beta_m n_m +O(1/N),
\l{x4e2}
$$
where classical energy is
$$
{\cal E}= H(\varphi^*,\varphi)=
\omega(K+1/2) - \frac{g}{4\omega^2} (K+1/2)^2.
$$

\subsection{Correspondence between asymptotic spectra}

Let us investigate the correspondence between found asymptotic formulas
for the spectrum, eqs.\r{x1} and \r{x4e2}. Let us set in eq.\r{x4e2}
$K=0$. In this case the frequency $\omega$ obeys the equation
$$
\omega^2 = a+ \frac{g}{2\omega}.
\l{xx1}
$$
Let us find the minimum of the potenrial \r{u1}. The corresponding
equation for $\overline{\xi}$ is
$$
-\frac{1}{4\overline{\xi}^3} + a\overline{\xi} + g\overline{\xi}^3 =0.
\l{xx2}
$$
However, eqs.\r{xx1} and \r{xx2} coincides if one sets:
$$
\overline{\xi}^2 = \frac{1}{2\omega}.
$$
It is easy to see that ${\cal E}=U(\overline{\xi})$, while
$\beta_2= \sqrt{U''(\overline{\xi})}$. Therefore, the
asymptotic spectrum in the case $K=0$ is
$$
E^n=NU(\overline{\xi}) + \varepsilon_0
+\sqrt{U''(\overline{\xi})}n_2 + \omega n_1 + 3\omega n_3+...
$$
The result can be interpretted as follows. There are many
perturbations around classical solution. However, the excitations with
energy $\beta_2$ do not break the quantum $O(N)$-symmetry, while other
excitations break this symmetry. This is a reason that they have not
been discovered by the collective-field approach which allows us to
find $O(N)$-symmetric solutions only. The method of second
quantization allows us also to find $O(N)$-asymmetric solutions.

We can also notice that one can perform a quantization around
solutions with $K\ne 0$. These solutions do not have form \r{a1w}, so
that the corresponding quantum states are $O(N)$-nonsymmetric. One can
say that quantization of the $K=0$-solution is investigation of
symmetric and nonsymmetric perturbations around $O(N)$-symmetric
quantum state while quantization around other solutions correspond to
research of the fluctuations around non-symmetric state. Thus, the
second-quantized approach allows us to find new energy levels of the
anharmonic oscillator.

\section{Formalism of    third    quantization    and    semiclassical
approximation}

This section deals with the third-quantized formulation of  the
quantum-field theory  \r{l}.  We  present  the Hamiltonian operator of
this theory   via   the   creation    and    annihilation    operators
$A^{\pm}(\varphi(\cdot))$. We  are  also going to consider the problem
of applicability   of   the   semiclassical    conception    to    the
third-quantized theory.     The     ``classical''     equations    for
$\Phi[\varphi(\cdot)]$ are to be derived.

\subsection{Creation and annihilation operators}

In the functional Schrodinger representation  states  of  the  N-field
system are  specified  by  (time-dependent) functionals \r{st},  while
their evolution is described by the functional Schrodinger equation
$$
i\frac{d\Psi_N}{dt} = H_N \Psi_N,
\l{ev1}
$$
where
$$
H_N = \int d{\bf x} \left( -\frac{1}{2}
\frac{\delta^2}{\delta\varphi^a({\bf x})\delta\varphi^a({\bf x})}+
\frac{1}{2}\nabla\varphi_a({\bf x})\nabla\varphi_a({\bf x})+
\frac{m^2}{2}\varphi_a({\bf x})\varphi_a({\bf x})\right)+
$$
$$
+\frac{\varepsilon\lambda}{4} \int d{\bf x} \varphi_a({\bf x})
\varphi_a({\bf x})\varphi_b({\bf x})\varphi_b({\bf x}),
$$
where $\varepsilon$ is a coupling constant in  the  $N$-field  theory,
$\varepsilon=1/N$.

The first  step  to  construct  the  third-quantized formulation is to
extend the Hilbert state space.  Instead of the space  of  functionals
$\Psi[\varphi_1(\cdot),...,\varphi_N(\cdot)]$, let   us  consider  the
``extended state space'' which is a Fock space of sets
$$
\left(
\begin{array}{c}
\Psi_0\\
\Psi_1[\varphi_1(\cdot)]\\
...\\
\Psi_N[\varphi_1(\cdot),...,\varphi_N(\cdot)]\\
...
\end{array}
\right)
\l{fock}
$$
where $\Psi_k$   is   symmetric  with  respect  to  transpositions  of
$\varphi_i$ and $\varphi_j$.  One could  conclude  that  the  physical
meaning of   $\Psi_0$,  $\Psi_1$,...,$\Psi_N$,...  is  the  following:
$\Psi_0$ is the  probability  amplitude  that  there  are  no  fields,
$\Psi_1$ is  the amplitude that there is only one field etc.  However,
theories with uncertain set of matter fields are not of widely use, so
that one should consider only such extended state vectors \r{fock} that
$\Psi_N\ne 0$ and  $\Psi_k=0$,  $k\ne  N$.  However,  considering  the
extended Schrodinger equation
$$
i\frac{d\Psi_k}{dt} = H_k \Psi_k, k=0,1,...
\l{ev2}
$$
on the Fock vector \r{fock} instead of eq. \r{ev1} will be shown to be
very useful,  since  system  of equations \r{ev2} is more suitable for
the semiclassical analysis.  One can construct asymptotic solution  to
the system  \r{ev2}  and  then  extract  the  $N$-th  component of the
extended state \r{fock}.

Eq.\r{ev2} can  be  presented  via  the  creation  and annihilation
operators. These  operators  $A^{\pm}(\varphi(\cdot))$  transform  the
extended state \r{fock} to the Fock vector
$$
\left(
\begin{array}{c}
(A^{\pm}[\varphi(\cdot)]\Psi)_0\\
(A^{\pm}[\varphi(\cdot)]\Psi)_1(\varphi_1(\cdot))\\
...\\
(A^{\pm}[\varphi(\cdot)]\Psi)_N(\varphi_1(\cdot),...,\varphi_N(\cdot))\\
...
\end{array}
\right),
\l{f1w}
$$
where
$$
(A^+[\varphi(\cdot)]\Psi)_k[\varphi_1(\cdot),...,\varphi_k(\cdot)]=
\frac{1}{\sqrt{k}}\sum_{a=1}^k \delta(\varphi(\cdot)-\varphi_a(\cdot))
\times
$$
$$
\times
\Psi_{k-1}[\varphi_1(\cdot),...,\varphi_{a-1}(\cdot),
\varphi_{a+1}(\cdot),...,\varphi_k(\cdot)],
$$
$$
\l{ca}
$$
$$
(A^-[\varphi(\cdot)]\Psi)_{k-1}
[\varphi_1(\cdot),...,\varphi_{k-1}(\cdot)]
= \sqrt{k} \Psi_k [\varphi(\cdot),
\varphi_1(\cdot),...,\varphi_{k-1}(\cdot)].
$$
The functional   $\delta$-function  entering  to  eq.  \r{ca}  can  be
formally defined via the functional integral over  all  spatial  field
configurations
$$
\int D\phi \delta(\phi(\cdot)-\varphi(\cdot))F[\phi(\cdot)]
=
F[\varphi(\cdot)]
$$
for arbitrary  functional $F$.  The formal integration measure $D\phi$
enters also to the inner product,
$$
(\Psi,\Psi)=\sum_{k=0}^{\infty} \int D\varphi_1... D\varphi_k
|\Psi[\varphi_1(\cdot),...,\varphi_k(\cdot)]|^2.
$$
The introduced  operators  \r{ca}  obey  usual  canonical  commutation
relations
$$
[A^{\pm}[\varphi(\cdot)], A^{\pm}[\phi(\cdot)] ] =0,
[A^-[\varphi(\cdot)], A^+[\phi(\cdot)]]=\delta(\varphi(\cdot)-
\phi(\cdot))
\l{ccr}
$$
Let us present different operators in the extended state space
\r{fock} through the operators \r{ca}.
It follows from eq.\r{ca} that the operator of ``number of
fields'' which multiplies the $k$-th component of \r{fock} by $k$ can
be presented as
$$
\int D\varphi A^+[\varphi(\cdot)] A^-[\varphi(\cdot)].
\l{n}
$$
Analogously, the operator
$$
\sum_{a=1}^k \varphi_a({\bf x}) \varphi_a({\bf y})
$$
can be written as
$$
\int D\varphi A^+[\varphi(\cdot)] \varphi({\bf x})
\varphi({\bf y}) A^-[\varphi(\cdot)].
\l{phi2}
$$
One has also the following relation
$$
\sum_{a=1}^k
\frac{\delta^2}{\delta \varphi_a({\bf x})
\delta \varphi_a({\bf y}) }
=
\int D\varphi A^+[\varphi(\cdot)]
\frac{\delta^2}{\delta \varphi({\bf x})
\delta \varphi({\bf y}) }
A^-[\varphi(\cdot)].
\l{p2}
$$
Making use of the obtained relations, we present the quantum
Hamiltonian via the creation and annihilation operators
$$
H= \int D\varphi A^+[\varphi(\cdot)] \int d{\bf x}
\left(
-\frac{1}{2}
\frac{\delta^2}{\delta \varphi({\bf x})
\delta \varphi({\bf x}) }+
\frac{1}{2}\nabla\varphi({\bf x})\nabla\varphi({\bf x})+
\frac{m^2}{2}\varphi({\bf x})\varphi({\bf x})\right)
A^-[\varphi(\cdot)]+
$$
$$
+
\frac{\varepsilon\lambda}{4} \int D\varphi D\phi
\int d{\bf x} \varphi^2({\bf x}) \phi^2({\bf x})
A^+[\varphi(\cdot)]A^-[\varphi(\cdot)]
A^+[\phi(\cdot)]A^-[\phi(\cdot)].
\l{h2w}
$$
The system of equations \r{ev2} can be presented in a simpler form
$$
i\frac{d\Psi}{dt}=H\Psi,
\l{m1}
$$
where $\Psi$ is an extended vector \r{fock}, $H$ is the operator
\r{h2w}.

\subsection{Perturbation theory or semiclassical approximation?}

Let us consider the case of small coupling constant $\varepsilon$. At
first sight, at small $\varepsilon$ the theory \r{h2w} can be
approximated by the theory of free fields. One can use then the
perturbation theory which can be shown to be equivalent to the
ordinary, second-quantized, perturbation theory.

However, we know from the quantum field theory that there are
non-perturbatve effects even at small values of coupling constant.
Namely, the conditions of applicability of such non-perturbative
methods as soliton quantization \cite{sol}, quantization in the
background of the strong external field \cite{str} or instanton
method \cite{tHooft,sol} is that the coupling constant
should be small.

To investigate the problem of applicability of
perturbative or non-perturbative
approach, consider the Heisenberg equations for the operators
$$
A^{\pm}_t[\varphi(\cdot)]=e^{iHt}A^{\pm}[\varphi(\cdot)]e^{-iHt}.
$$
which have the form
$$
i\frac{d}{dt}A^{\pm}_t[\varphi(\cdot)]=
\pm [A^{\pm}_t[\varphi(\cdot)],H]
$$
and can be simplified by using the canonical commutation relations
\r{ccr}. The equation for the annihilation operator is
$$
i\frac{d}{dt}A^{-}_t[\varphi(\cdot)]=
\int d{\bf x}
\left(
-\frac{1}{2}
\frac{\delta^2}{\delta \varphi({\bf x})
\delta \varphi({\bf x}) }+
\frac{1}{2}\nabla\varphi({\bf x})\nabla\varphi({\bf x})+
\frac{m^2}{2}\varphi({\bf x})\varphi({\bf x})\right)
A^-_t[\varphi(\cdot)]+
$$
$$
+
\frac{\varepsilon\lambda}{2}  \int D\phi
\int d{\bf x} \varphi^2({\bf x}) \phi^2({\bf x})
A^+_t[\phi(\cdot)]A^-_t[\phi(\cdot)]
A^-_t[\varphi(\cdot)]
\l{hs1}
$$
the conjugated equation is the equation for the creation operator.

The usual procedure to derive the classical equations of motion is
averaging of eq.\r{hs1}. If the second term of the right-hand side of
eq.\r{hs1} can be neglected, the perturbation theory is applicable.
Otherwise, one should develop a non-perturbative approach. The
difficulty may arise from the factor
$$
\int D\phi A^+[\phi(\cdot)]\phi^2({\bf x}) A^-[\phi(\cdot)]
$$
which is of order $k$ when it is applied to the $k$-field component.
Thus, we see that if the number of fields $N$ satisfies the condition
$$
N << 1/\varepsilon
$$
the perturbation theory is applicable, while if
$$
N \sim 1/\varepsilon,
$$
another approach is necessary.

\subsection{Classical equations}

\subsubsection{Averaging the Heisenberg equations}

One of the approaches to the semiclassical theory is
the following. Let us rescale the creation and annihilation
operators
$$
\Phi^{\pm}[\varphi(\cdot)] = \sqrt{\varepsilon}
A^{\pm}[\varphi(\cdot)].
$$
The Heisenberg equation \r{hs1} will take the form
$$
i\frac{d}{dt}\Phi^{-}_t(\varphi(\cdot))=
\int d{\bf x}
\left(
-\frac{1}{2}
\frac{\delta^2}{\delta \varphi({\bf x})
\delta \varphi({\bf x}) }+
\frac{1}{2}\nabla\varphi({\bf x})\nabla\varphi({\bf x})+
\frac{m^2}{2}\varphi({\bf x})\varphi({\bf x})\right)
\Phi^-_t(\varphi(\cdot))+
$$
$$
+
\frac{\lambda}{2}  \int D\phi
\int d{\bf x} \varphi^2({\bf x}) \phi^2({\bf x})
\Phi^+_t(\phi(\cdot))\Phi^-_t(\phi(\cdot))
\Phi^-_t(\varphi(\cdot))
\l{hs2}
$$
which does not contain the small parameter $\varepsilon$.

The coupling constant $\varepsilon$ appears, however, in the canonical
commutation relations
$$
[\Phi^-_t[\varphi(\cdot)], \Phi^+_t[\phi(\cdot)]]=
\varepsilon \delta(\phi(\cdot)-\varphi(\cdot)).
\l{ccr3}
$$
The fact that the small parameter arises in the commutatora and does
not arise in the dynamical equations resembles semiclassical quantum
mechanics: the Heisenberg equations are regular in semiclassical limit
(they transforms to Hamiltonian system), while the commutator between
coordinate and momenta operator tends to zero.

Analogously, one can try to substitute the operators entering to
eq.\r{hs2} by the $c$-number quantities,
$$
\Phi_t^- \to \Phi_t, \qquad \Phi_t^+ \to \Phi_t^*,
$$
and obtain the classical equation
$$
i\frac{d}{dt}\Phi_t[\varphi(\cdot)]=
\int d{\bf x}
\left(
-\frac{1}{2}
\frac{\delta^2}{\delta \varphi({\bf x})
\delta \varphi({\bf x}) }+
\frac{1}{2}\nabla\varphi({\bf x})\nabla\varphi({\bf x})+
\frac{m^2}{2}\varphi({\bf x})\varphi({\bf x})\right)
\Phi_t[\varphi(\cdot)]+
$$
$$
+
\frac{\lambda}{2}  \int D\phi
\int d{\bf x} \varphi^2({\bf x}) \phi^2({\bf x})
\Phi_t^*(\phi(\cdot))\Phi_t(\phi(\cdot))
\Phi_t(\varphi(\cdot))
\l{hs3}
$$
Such substitution can be explained via the averaging procedure of
eq.\r{hs2} analogously to the derivation of the Ehrenfest theorem in
quantum mechanics. Since the commutator between operators
$\Phi^{\pm}_t$ is small, the uncertainty of them can be made of order
$O(\sqrt{\varepsilon})$. This means that for some states average value
$$
<A(\Phi^+_t,\Phi^-_t)>
\l{a2z}
$$
can be approximated up to $O(\sqrt{\varepsilon})$ by the quantity
$A(\Phi^*_t,\Phi_t)$,
$$
<A(\Phi^+_t,\Phi^-_t)> \to
A(\Phi^*_t,\Phi_t).
\l{a1}
$$
This means that classical limit of eq.\r{hs2} is
eq.\r{hs3}.

The states obeying eq.\r{a1} can be constructed as follows. Let $Y$ be
$\varepsilon$-independent extended state \r{fock}. Consider the
unitary operator
$$
U_{\Phi}= \exp\left[
\frac{1}{\sqrt{\varepsilon}}\int D\phi
(\Phi(\varphi(\cdot))A^+(\varphi(\cdot))
-\Phi^*(\varphi(\cdot))A^-(\varphi(\cdot)))
\right]
\l{uf}
$$
obeying the relations
$$
\begin{array}{c}
U_{\Phi}^{-1} A^- U_{\Phi} = A^- + \frac{1}{\sqrt{\varepsilon}} \Phi
\\
U_{\Phi}^{-1} A^+ U_{\Phi} = A^+ + \frac{1}{\sqrt{\varepsilon}} \Phi^*
\end{array}
\l{cr2w}
$$
which are corollaries of the canonical commutation relations \r{ccr}.
Consider the semiclassical state
$$
U_{\Phi}Y.
\l{semicl}
$$
It follows from eqs.\r{cr2w} that average value \r{a2z} is equal to the
average value of the operator
$
A(\Phi^*_t+\sqrt{\varepsilon} A^+_t, \Phi_t+ \sqrt{\varepsilon} A^-_t)
$
over $\varepsilon$-independent state $Y$. Making a limit
$\varepsilon\to 0$, we obtain eq.\r{a1}.

Thus, we see that semiclassical analysis of the $N$-field equation
based on the Heisenberg approach may lead us to classical equation
\r{hs3}. The classical variable is a complex fiunctional
$\Phi[\varphi(\cdot)]$ being the state of quantum one-field
configuration. This state obeys unusual nonlinear equation.

The ``semiclassical'' state \r{semicl}
is a Fock vector \r{fock} with all non-zero components. One can try to
extract from eq.\r{semicl} one of the component, for example, the
$N$-th component. However, one should be careful because our
calculations are not exact but approximate. If the $N$-th component of
eq.\r{semicl} is exponentially small as $\varepsilon\to 0$,it should
be neglected, so that the approximate wave functional will vanish, and
 the constructed asymptotics will be trivial. For the simplest
case ($Y$ is a vacuum) the probability that there are $N$ fields is
given by the Poisson distribution with the maximum at
$$
N= \frac{1}{\varepsilon} \int D\phi |\Phi(\phi(\cdot))|^2.
\l{nmax}
$$
This relation can be derived also from eqs.\r{n} and
\r{a1}. Eq.\r{nmax} means that semiclassical approximation allows us
to investigate the large-N field theory if $N\sim 1/\varepsilon$.
Note that the case $N\sim 1$ corresponds to
$\Phi\sim\sqrt{\varepsilon}$, classical equation \r{hs3} transforms to
the equation for free fields, so that the perturbation theory is
applicable.

\subsubsection{BBGKY-like approach}

Eq.\r{hs3} can be also derived by the BBGKY (Bogoliubov - Born - Green
- Kirkwood - Yvon) approach (\cite{B}, see also \cite{BM})
from the $N$-field equation \r{ev1}. Let us briefly discuss this
technique.

Consider the $k$-field correlation functional
$$
R_k^t(\varphi_1(\cdot),...,\varphi_k(\cdot);\phi_1(\cdot),...,
\phi_k(\cdot))=
\int D\varphi_{k+1}... D\varphi_{N}
$$
$$
\times
\Psi^*_N[\varphi_1(\cdot),
...,\varphi_N(\cdot)]          \Psi_N[\phi_1(\cdot),...,\phi_k(\cdot),
\varphi_{k+1}(\cdot),...,\varphi_N(\cdot)]
\l{cor}
$$
corresponding to  the  $N$-field   wave   functional   $\Psi_N$.   The
correlator \r{cor} can be presented via the creation and annihilation
operators
$$
\frac{(N-k)!}{N!} <A^+[\varphi_1(\cdot)]...A^+[\varphi_k(\cdot)]
A^-[\phi_1(\cdot)]...A^-[\phi_k(\cdot)]>.
$$
It follows from eq.\r{hs1} that functionals \r{cor} obey the
BBGKY-like hierarchy of equations.

Suppose that the correlators \r{cor} factorize as $k=const$,
$N\to\infty$
$$
R^t_k \sim \Phi_t^*[\varphi_1(\cdot)]...\Phi_t^*[\varphi_k(\cdot)]
\Phi_t[\phi_1(\cdot)]...\Phi_t[\phi_1(\cdot)].
\l{c1}
$$
Substituting eq.\r{c1} to the BBGKY-hierarchy, one can find that for
some phase factor $e^{i\gamma_t}$ the functional $\Phi_te^{i\gamma_t}$
obeys eq.\r{hs3}.

\subsubsection{Variational principles}

The classical equation of motion \r{hs3} can be presented as a
Hamiltonian system. Namely, this equation is derivable from the
variational principle
$$
\int_0^t dt \left[\int D\varphi \Phi_t^*[\varphi(\cdot)] i\frac{d}{dt}
\Phi_t[\varphi(\cdot)] \right. -
$$
$$
\int D\varphi \Phi^*_t[\varphi(\cdot)] \int d{\bf x}
\left(
-\frac{1}{2}
\frac{\delta^2}{\delta \varphi({\bf x})
\delta \varphi({\bf x}) }+
\frac{1}{2}\nabla\varphi({\bf x})\nabla\varphi({\bf x})+
\frac{m^2}{2}\varphi({\bf x})\varphi({\bf x})\right)
\Phi_t[\varphi(\cdot)]-
$$
$$
-
\frac{\lambda}{4} \int D\varphi D\phi
\int d{\bf x} \varphi^2({\bf x}) \phi^2({\bf x})
|\Phi_t[\varphi(\cdot)]|^2|\Phi_t[\phi(\cdot)]|^2
\left.\right] \to extr.
\l{var}
$$
This variational principle is Hamiltonian. Namely, one can extract
real and imaginary part
$$
\Phi=\frac{Q+iP}{\sqrt{2}}
$$
and notice that the principle \r{var} is
$$
\int dt (P\dot{Q}-H) \to extr.
$$
Thus, classical equations of motion correspond to infinite-dimensional
flat phase space rather than to curved space considered in
ref.\cite{BY,CM5}.

We will see that the (time-dependent) Gaussian ansatz
$$
\Phi[\varphi(\cdot)]= c \exp\left(
\frac{i}{2} \int d{\bf x} d{\bf y} \varphi({\bf x})
A({\bf x},{\bf y})\varphi({\bf y}))
\right)
\l{g}
$$
satisfies eq.\r{hs3}, so that one can consider the subspace of the
phase space which is associated with Gaussian states \r{g}. The
corresponding variational principle for classical mechanics can be
obtained by substitution eq.\r{g} to eq.\r{var}. The obtained dynamics
coincides with refs.\cite{BY,CM5}.

However, one can consider other solutions to eq.\r{hs3}, for example,
the product of the polynomial in $\varphi$ by the Gaussian exponent
\r{g}. These solutions will not belong to the reduced curved phase
space but will belong to the flat space $\{\Phi[\varphi(\cdot)]\}$.

The variational principle \r{var} can be interpretted as follows. One
can consider the time-dependent variational principle \cite{JK,GR}
$$
\int_0^t dt (\psi_N^t, (i\frac{d}{dt}-H_N) \psi_N^t)
\l{v2}
$$
for the $N$-field equation and substitute the $N$-particle
time-dependent test functional
$$
\Psi_N^t[\varphi_1(\cdot),...,\varphi_N(\cdot)]=
\Phi_t[\varphi_1(\cdot)]...\Phi_t[\varphi_N(\cdot)]
\l{pr}
$$
to eq.\r{v2}. One obtains then the classical variational principle
\r{var}.

Note that the Gaussian ansatz to the quantum variational principle
instead of \r{pr} lead to the classical mechanics in the reducrd
curved space.

Although the wave functional \r{pr} will be shown to be {\it not} the
asymptotic solution to eq.\r{ev1}, eq.\r{hs3} is correct and can be
derived also by the substituting the asymptotics for the wave
functional to eq.\r{ev1}. This will be done in section 4.

\subsubsection{Operator formulation of classical mechanics}

Let us discuss relation between eq.\r{hs3} and Heisenberg approach
\cite{CM4}. Consider the ``Heisenberg'' field
$\hat{\varphi}^t({\bf x})$ which
can be defined as follows. Let $U^t$ be a {\it non-linear} operator
transfering initial condition for eq.\r{hs3} to the solution to
eq.\r{hs3} at time t:
$$
\Phi_t=U_t(\Phi_0).
$$
The Heisenberg operator is
$$
\hat{\varphi}^t({\bf x})=U_t^{-1}\varphi({\bf x})U_t
$$
such that
$$
(\Phi,\hat{\varphi}^t({\bf x})\Phi)=
(\Phi^t,\hat{\varphi}({\bf x})\Phi^t).
\l{hs4}
$$
It follows from eqs.\r{hs3} and \r{hs4} that the Heisenberg field
$\hat{\varphi}^t({\bf x})=\hat{\varphi}(x)$ obeys the equation
$$
\partial_{\mu}\partial_{\mu}\hat{\varphi}(x)+
m^2\hat{\varphi}(x)+ \lambda <\hat{\varphi}^2(x)>\hat{\varphi}(x)
=0
\l{hs5}
$$
which was obtained in \cite{CM4}.

An interesting feature of this second-quatized theory which is
classical for the $N$-field model is that the field
$\hat{\varphi}(x)$ is a non-linear operator in the one-field state
space. We will discuss in the next sections the applications of the
semiclassical approach to eq.\r{ev1} and clarify the role of
eq.\r{hs3} in constructing asymptotic solutions to the $N$-field
equation.

\subsection{Regularization and renormalization}

In the  previous  subsection  we  have  derived the classical equation
\r{hs3} on the functional $\Phi_t[\varphi(\cdot)]$. However, this
derivation was formal. We have not taken into account the problem of
divergences and renormalization in quantum field theory. Thus,
eq.\r{hs3} is not well-defined, and it is necessary to investigate the
problem of correct definition of the classical equations in details.

\subsubsection{Regularization}

There are many ways to regularize the quantum field theory.
Relativistic-invariant regularizations are usually applied to
evaluation of Feynman graphs, while lattice regularization is studied
in non-perturbative approaches. Another way to regularize the theory
is to substitute the field $\varphi({\bf x})$ by the cutoffed field
$\varphi_{\Lambda}({\bf x})$:
$$
 \varphi_{\Lambda}({\bf x}) = \int d{\bf y}A_{\Lambda}({\bf x}-{\bf y})
\varphi({\bf y}),
\l{z1}
$$
where $A_{\Lambda}({\bf x}-{\bf y}) \to
\delta({\bf x}-{\bf y})$ as the parameter of the ultraviolet cutoff
$\Lambda$ tends to infinity. Eq.\r{z1} allows us to regularize the
canonical commutation relations between field and momentum:
$$
[ \varphi_{\Lambda}({\bf x}), \pi({\bf x})] =
i A_{\Lambda}({\bf x}-{\bf y})
$$
instead of
$$
[ \varphi({\bf x}), \pi({\bf x})] =
i \delta({\bf x}-{\bf y}).
$$
To perform infrared regularization, we consider the theory in the box
with sizes $L\times L \times ...\times L$ with the periodic boundary
conditions.

After regularization eq.\r{hs3} is written as
$$
i\dot{\Phi}_t[\varphi(\cdot)] =
\int d{\bf x}
\left(\frac{1}{2} \pi^2({\bf x}) + \frac{1}{2} (\nabla \varphi_{\Lambda})^2
({\bf x}) + \frac{u_t({\bf x})}{2}\varphi_{\Lambda}^2({\bf x})\right)
{\Phi}_t[\varphi(\cdot)],
\l{z2}
$$
where
$$
u_t({\bf x})=m^2 + \lambda (\Phi_t,\varphi_{\Lambda}^2({\bf x})\Phi_t),
(\Phi_t,\Phi_t)=1.
\l{z3}
$$
Let us investigate the system of equations \r{z2},\r{z3} and consider
the divergences in this system.

\subsubsection{Renormalization in the Schrodinger picture}

Consider the problem of renormalizing the obtained equations.
For the  spatially homogeneous classical field,  analogous problem was
considered in \cite{GR,B4},  while some examples for the inhomogeneous
field case are presented in \cite{B5}. Another approach
was considered in refs. \cite{MS7,MS8}. The idea is that one
should impose the conditions not only on the counterterms entering to
the Hamiltonian but also on the state vector. One cannot consider the
state vectors which are regular as the cutoffs tends to infinity. One
should specify how the state vector depends on the parameters of the
cutoffs; this dependence is singular. Then one should check that these
conditions are invariant under time evolutions.

Let us perform the renormalization procedure in such a way for
Gaussian and non-Gaussian solutions to eq.\r{z2}.

\subsubsection{Gaussian and non-Gaussian states}

We are going to investigate eq.\r{z2}.
First of all, suppose that the function $u_t({\bf x})$ is known. We
will require this function to be non-singular:
$$
\lim_{\Lambda\to\infty} u_t({\bf x}) < \infty
\l{z4}
$$
because it enters to classical equations of motion for the model
\r{z2}.

Consider the wave functional being a product of a polynomial by the
Gaussian exponent:
$$
\sum_n \int d{\bf x}_1...d{\bf x}_n f_n({\bf x}_1,...,{\bf x}_n)
\varphi({\bf x}_1)...\varphi({\bf x}_n)
\exp\left[\frac{i}{2}\int d{\bf x} d{\bf y} \varphi({\bf x})R({\bf x},{\bf
y})\varphi({\bf y})\right].
\l{z5}
$$
To investigate the divergences in the solution of the Cauchy problem
for eq.\r{z2}, it is convenient to notice that the operator
$$
B^+[P_t,Q_t]=
\int d{\bf x} \left[\varphi({\bf x})P_t({\bf x})-
\frac{1}{i}\frac{\delta}{\delta \varphi({\bf x})} Q_t({\bf x})\right]
\l{z6}
$$
commutes with the operator $i\frac{d}{dt}-H$ ($H$ is the operator
entering to the right-hand side of eq.\r{z2}) if
$$
\dot{Q}_t({\bf x})=P_t({\bf x}),
$$
$$
\l{z7}
$$
$$
\dot{P}_t({\bf x}) + \int d{\bf y}d{\bf z}A_{\Lambda}({\bf x}-{\bf y})
(-\Delta_{\bf y} +u_t({\bf y}))
A_{\Lambda}({\bf y}-{\bf z})Q_t({\bf z})=0.
$$
The initial condition \r{z5} can be presented as a linear combination
of the states like:
$$
B^+[P^1_0,Q^1_0] ... B^+[P^m_0,Q^m_0] \Phi,
\l{z8}
$$
where $\Phi$ is a Gaussian state. The solution to eq.\r{z2} which obeys
the initial condition \r{z8} is expressed via the solutions
$(P^1_t,Q^1_t)$, ..., $(P^m_t,Q^m_t)$ to the system eq. \r{z7},
$$
B^+[P^1_t,Q^1_t] ... B^+[P^m_t,Q^m_t] \Phi^t.
$$
Since eqs.\r{z7} have regular limits as $\Lambda\to\infty$, the
operators \r{z6} are regular as $\Lambda\to \infty$. This means that
singularities may arise in the Gaussian state $\Phi^t$ only.

Let us consider the Gaussian solution to eq.\r{z2}:
$$
c_t
\exp\left[\frac{i}{2}\int d{\bf x} d{\bf y} \varphi({\bf x})
R_t({\bf x},{\bf y})\varphi({\bf y})\right].
\l{z9}
$$
Substitute formula \r{z9} to the Schrodinger equation \r{z2}. By
$\hat{R}$ we denote the operator with kernel
$R({\bf x},{\bf y})$:
$$
(\hat{R}f)({\bf x})=\int d{\bf y}R({\bf x},{\bf y}) f({\bf y}).
$$
Analogously, by $\hat{A}$ we denote the operator with kernel
$A_{\Lambda}({\bf x},{\bf y})$. We find that the Gaussian quadratic
form obeys the nonlinear equation:
$$
\frac{d}{dt}\hat{R}_t + \hat{R}_t \hat{R}_t
+ \hat{A} (-\Delta + u_t) \hat{A} =0,
\l{z10}
$$
where $u_t$ is the operator of multiplication by $u_t({\bf x})$. The
prefactor $c_t$ satisfies the equation
$$
\frac{d}{dt}(ln c_t)=-\frac{1}{2} Tr \hat{R}_t.
\l{z11}
$$

\subsubsection{Operational calculus}

To investigate singularities in eq.\r{z10}, it is convenient
\cite{M4,KM} to
consider the symbols of the operators, i.e. to present each operator
via operators ${\bf x}$ and $-i\frac{\partial}{\partial {\bf x}}$, for
example:
$$
\hat{R}_t=R_t(\stackrel{2}{\bf x},
\stackrel{1}{-i\frac{\partial}{\partial {\bf x}}}).
\l{z12}
$$

Since the operators  ${\bf x}$ and $-i\frac{\partial}{\partial {\bf x}}$
do not commute,  it is important to specify ordering of the operators.
Notation \r{z12}  means  that  momenta   operators   act   first   and
multiplication operators  act  next,  so  that  the  operator  \r{z12}
transforms the vector
$$
f_{{\bf p}}({\bf x})= const e^{i{\bf p}{\bf x}}
$$
to the vector
$$
(\hat{R}_tf_{\bf p})({\bf x}) = R_t({\bf x},{\bf p})f_{\bf p}({\bf x})
\l{z13}
$$
Eq.\r{z13} can  be  treated  as  a  definition  of  the  symbol of the
operator $\hat{R}_t$.

Let us consider the product of the operators
$
\hat{B}=B(\stackrel{2}{\bf x},
\stackrel{1}{-i\frac{\partial}{\partial {\bf x}}})
$
and
$
\hat{C}=C(\stackrel{2}{\bf x},
\stackrel{1}{-i\frac{\partial}{\partial {\bf x}}})
$.
It can be also presented via coordinate and momentum operators:
$$
\hat{B}\hat{C}=(B*C)(\stackrel{2}{\bf x},
\stackrel{1}{-i\frac{\partial}{\partial {\bf x}}}).
$$
It happens that the ``product'' $B*C$ can be presented as
$$
(B*C)({\bf x},{\bf k})=
B(\stackrel{2}{\bf x},
\stackrel{1}{{\bf k}-i\frac{\partial}{\partial {\bf x}}})
C({\bf x},{\bf k}).
\l{z14}
$$
To prove formula \r{z14}, one can consider first the case
$$
C({\bf x},{\bf k})=e^{i{\bf p}{\bf x}}C({\bf k}).
\l{z15}
$$
In this case
$$
\hat{C}f_{\bf k}({\bf x})=C({\bf k})f_{{\bf k}+{\bf p}}({\bf x}),
$$
$$
\hat{B}\hat{C}f_{\bf k}({\bf x})=B({\bf x},{\bf k}+{\bf p})
C({\bf k})f_{{\bf k}+{\bf p}}({\bf x})=
B({\bf x},{\bf k}+{\bf p})
e^{i{\bf p}{\bf x}}C({\bf k})
f_{\bf k}({\bf x}),
$$
so that
$$
(B*C)({\bf x},{\bf k})=
B({\bf x},{\bf k}+{\bf p})
C({\bf x},{\bf k}),
$$
$$
B(\stackrel{2}{\bf x},
\stackrel{1}{{\bf k}-i\frac{\partial}{\partial {\bf x}}})
C({\bf x},{\bf k}) =
B({\bf x},{\bf k}+{\bf p})
C({\bf x},{\bf k})
$$
Formula \r{z14} is then proved for this partial  case.  To  check  eq.
\r{z14} for  general  case,  notice  that  if  eq.\r{z14} is valid for
operators $\hat{C}=\hat{C}_1$  and  $\hat{C}=\hat{C}_2$,  it  is  also
valid for  $\hat{C}=\hat{C}_1  +  \hat{C}_2$.  But any function can be
presented as a linear  combination  of  functions  \r{z15}.  Thus,  we
justify eq.\r{z14} for general case.

Eq.\r{z10} can  be  considered  as  an  equation for the symbol of the
operator $R_t$:
$$
\dot{R}_t({\bf x},{\bf k})+ (R_t*R_t)({\bf x},{\bf k}) +
A_{\bf k}*({\bf k}^2+u_t({\bf x})) A_{\bf k}=0.
\l{z16}
$$
We have  taken  into  account  that  the  operator  $\hat{A}$  can  be
presented as  $A(-i\frac{\partial}{\partial {\bf x}})$,  where $A({\bf
k})=A_{\bf k}$  is  a   Fourier   transformation   of   the   function
$A_{\Lambda}$. As $\Lambda \to\infty$, $A_{\bf k} \to 1$.

\subsubsection{Singularities of the Gaussian quadratic form}

To investigate  the  ultraviolet  singularities,  it  is  important to
consider the behaviour of the symbol of the operator $R_t$ as ${\bf k}
\to \infty$.  It is convenient to expand eq.\r{z14} into an asymptotic
series:
$$
(B*C)({\bf x},{\bf k})=
\sum_{l\ge 0} \frac{(-i)^l}{l!}
\frac{\partial^l B({\bf x},{\bf k})}
{\partial k_{i_1}...\partial k_{i_l}}
\frac{\partial^l C({\bf x},{\bf k})}
{\partial x_{i_1}...\partial x_{i_l}}.
\l{z15a}
$$
This is an expansion in $1/|{\bf k}|$,  because the ratio between  the
next order of the series and the previous one is
$O(1/|{\bf k}|)$. We see that in the leading order of
$1/|{\bf k}|$
$$
(B*C)({\bf x},{\bf k}) \sim B({\bf x},{\bf k})
C({\bf x},{\bf k}).
$$
Let us consider the behaviour of eq.\r{z16} as ${\bf k}\to\infty$.
Since the  third  term of the left-hand side of eq.\r{z16} is of order
$|{\bf k}|^2$, one can think that $R_t \sim |{\bf k}|$ as $|{\bf k}|\to
\infty$. Thus, it is reasonable to extract the most singular term from
$R_t$:
$$
R_t({\bf x},{\bf k}) = iA_{\bf k}\sqrt{{\bf k}^2+\mu^2}
+r_t({\bf x},{\bf k}).
\l{z16a}
$$
Eq.\r{z16} takes the form
$$
\dot{r}_t({\bf x},{\bf k}) + 2i A_{\bf k} \sqrt{{\bf k}^2+\mu^2}
r_t({\bf x},{\bf k}) + A_{\bf k}(u_t({\bf x})-\mu^2)A_{\bf k}=
O(|{\bf k}|^{-1}).
$$
We see that the quantity $r_t({\bf x},{\bf k})$ should be of order
$|{\bf k}|^{-1}$:
$$
r_t({\bf x},{\bf k})\sim \frac{iA_{\bf k}(u_t({\bf x})-\mu^2)}
{2\sqrt{{\bf k}^2+\mu^2}} + O(|{\bf k}|^{-2}).
\l{z16b}
$$
Analogously, one  can find a solution to the equation \r{z10} with the
help of the iteration procedure up to $O(|{\bf k}|^{-\alpha})$
for arbitrary $\alpha$.

Let us impose at the initial time moment the condition
$$
R_t({\bf x},{\bf k}) = R_t^{sing}({\bf x},{\bf k})
+R_t^{reg}({\bf x},{\bf k}),
\l{z17}
$$
where $R_t^{sing}$  is  the  obtained  solution  to eq.\r{z10},  while
$R_t^{reg}$ is of order $O(|{\bf k}|^{-\alpha})$. Then eq.\r{z17} will
be invariant  under  time  evolution.  One  can therefore find all the
singularities of the quadratic form entering to the Gaussian exponent.

\subsubsection{Renormalization of mass and coupling constant}

We are now going to check  our  main  condition  \r{z4}.  It  is  then
necessary to find the singularities of the average value of
$\varphi^2_{\Lambda}({\bf x})$.
This matrix element can be written via the
functional integral
$$
<\varphi^2_{\Lambda}({\bf x})>=
\int D\varphi P[\varphi(\cdot)]
\varphi^2_{\Lambda}({\bf x}) \exp[-(\varphi,Im \hat{R} \varphi)],
\l{z18}
$$
where $P$ is a polynomial functional of $\varphi(\cdot)$ with smooth
coefficient functions. Calculation of this integral is standard: it is
presented as
$$
P\left[\frac{\delta}{\delta j(\cdot)}\right]
\int d{\bf y} d{\bf z} A_{\Lambda}({\bf x}-{\bf y})
 A_{\Lambda}({\bf x}-{\bf z})
\frac{\delta^2}{\delta j({\bf y})\delta j({\bf z})}
\int D\varphi e^{(j,\varphi)-(\varphi,Im \hat{R} \varphi)},
$$
Since the remaining integral is easily evaluated, the latter
expression can be simplified:
$$
P\left[\frac{\delta}{\delta j(\cdot)}\right]
\int d{\bf y} d{\bf z} A_{\Lambda}({\bf x}-{\bf y})
 A_{\Lambda}({\bf x}-{\bf z})
\frac{\delta^2}{\delta j({\bf y}\delta j({\bf z})}
e^{\frac{1}{4}(j,(Im \hat{R})^{-1}j)}.
\l{z19}
$$
The only singularity in this average value comes from the operator
$\frac{\delta^2}{\delta j\delta j}$. This means that the matrix
element consists of the regular and singular parts:
$$
<\varphi^2_{\Lambda}({\bf x})>=
\frac{1}{2}
<{\bf x}|\hat{A}(Im \hat{R})^{-1}\hat{A}|{\bf x}>+
<\varphi^2_{\Lambda}({\bf x})>_{reg}.
$$
Since the symbol of the operator $R_t$ is
$$
R_t({\bf x},{\bf k}) = iA_{{\bf k}}\sqrt{{\bf k}^2+\mu^2}
\left(1+\frac{u_t({\bf x})-\mu^2}{2({{\bf k}^2+\mu^2})}+
O(|{\bf k}|^{-3})\right),
$$
the leading orders for the symbol of the operator
$\hat{A}(Im \hat{R})^{-1}\hat{A}$ have the following form:
$$
(\hat{A}(Im \hat{R})^{-1}\hat{A})({\bf x},{\bf k})=
\frac{ A_{{\bf k}}}{\sqrt{{\bf k}^2+\mu^2}}
(1-\frac{u_t({\bf x})-\mu^2}{2({{\bf k}^2+\mu^2})}+
O(|{\bf k}|^{-3})).
$$
Since the matrix element of arbitrary operator $\hat{B}$ is expressed
via its symbol as
$$
<{\bf x}|\hat{B}|{\bf x}>=
\sum_{\bf p} |<{\bf x}|{\bf p}>|^2 B({\bf x},{\bf p})=
\frac{1}{L^d} \sum_{\bf p} B({\bf x},{\bf p}),
$$
the average value \r{z18} is presented up to a finite quantity
$<\varphi^2({\bf x})>_{reg}$ in the following form:
$$
<\varphi^2_{\Lambda}({\bf x})>= \frac{1}{2L^d}
\sum_{\bf k} \frac{A_{\bf k}}{\sqrt{{\bf k}^2+\mu^2}} -
\frac{u_t({\bf x})-\mu^2}{4L^d} \sum_{\bf k}
\frac{A_{\bf k}}{({\bf k}^2+\mu^2)^{3/2}}
+<\varphi^2({\bf x})>_{reg}.
\l{z21}
$$
Substituting eq.\r{z21} to eq.\r{z3}, one obtains the following
condition:
$$
m^2+ \lambda \left[
\frac{1}{2L^d}
\sum_{\bf k} \frac{A_{\bf k}}{\sqrt{{\bf k}^2+\mu^2}} -
\frac{u_t({\bf x})-\mu^2}{4L^d} \sum_{\bf k}
\frac{A_{\bf k}}{({\bf k}^2+\mu^2)^{3/2}}
\right] + \lambda
<\varphi^2({\bf x})>_{reg} = u_t({\bf x}).
\l{z22}
$$
Let us investigate eq.\r{z22}. First of all, set
$$
\mu^2=m^2+\frac{\lambda}{2L^d} \sum_{\bf k}
\frac{A_{\bf k}}{\sqrt{{\bf k}^2+\mu^2}}.
\l{z23}
$$
Since the sum in the right-hand side of eq.\r{z23} diverges as
$\Lambda\to\infty$, one should choose the quantity $m^2$ (square of
the bare mass) to be infinite to make $\mu^2$ finite. We will see in
the following sections that the quantity $\mu$ is a physical mass of
elementary particles, so that it must be finite.

Eq.\r{z22} can be presented then as
$$
\frac{u_t({\bf x})-\mu^2}{\lambda}=
<\varphi^2({\bf x})>_{reg} -
\frac{u_t({\bf x})-\mu^2}{L^d}
\sum_{\bf k}
\frac{A_{\bf k}}{4({\bf k}^2+\mu^2)^{3/2}}.
$$
The sum over ${\bf k}$ entering to this equation is infinite for the
(3+1)-dimensional case. One should then choose the bare coupling
constant $\lambda$ in such a way that
$$
\frac{1}{\lambda}
+\frac{1}{L^d}
\sum_{\bf k}
\frac{A_{\bf k}}{4({\bf k}^2+\mu^2)^{3/2}}
=\frac{1}{\lambda_R} < \infty.
\l{z24}
$$
The quantity $\lambda_R$ is a renormalized coupling constant.

\subsection{Complex-WKB approach for the third-quantized systems}

In subsection  3.3 we have found the classical equations with the help
of conjecture \r{a1} on the average values of the Heisenberg
operators. Let us now check this conjecture and develop a systematic
semiclassical theory. We will substitute an ansatz for the
third-quantized state to the equation of motion. We show that the
constructed state approximately satiisfies the equation.

\subsubsection{The complex-WKB ansatz}

Let us consider the following vector of the third-quantized Fock space:
$$
\Psi=\exp\left(\frac{i}{\varepsilon}S^t\right)U_{\Phi^t} Y^t,
\l{m2}
$$
where $S^t$ is a real number, $U_{\Phi^t}$ is the unitary operator
\r{uf}, $Y^t$ is an $\varepsilon$-independent state. It happens that
the third-quantized state \r{m2} being a vector \r{fock} approximately
obeys the equation \r{m1}:
$$
i\frac{d\Psi}{dt} = \frac{1}{\varepsilon}
H(\sqrt{\varepsilon}A^+, \sqrt{\varepsilon}A^-) \Psi.
\l{m3}
$$
Namely, one can use the commutation relations \r{cr2w}, as well as the
formula
$$
U_{\Phi^t}^{-1} \frac{d}{dt} U_{\Phi^t} =
$$
$$
=\frac{1}{\sqrt{\varepsilon}}
\int D\varphi (\dot{\Phi}^t[\varphi(\cdot)] A^+[\varphi(\cdot)]
-\dot{\Phi}^{t*}[\varphi(\cdot)] A^-[\varphi(\cdot)]  )+
\frac{1}{2\varepsilon} \int D\varphi (
{\Phi}^{t*}[\varphi(\cdot)]\dot{\Phi}^{t}[\varphi(\cdot)]
-
\dot{\Phi}^{t*}[\varphi(\cdot)]{\Phi}^{t}[\varphi(\cdot)]).
$$
Analogously to appendix B, one can make equal first the terms of order
$O(1/\varepsilon)$ in eq.\r{m3}:
$$
\frac{dS^t}{dt} = -\frac{i}{2\varepsilon} [(\dot{\Phi}^t,\Phi^t)
- ({\Phi}^t,\dot{\Phi}^t)] -H(\Phi^*,\Phi).
\l{m4}
$$
Then one can consider the terms of order $O({\varepsilon}^{-1/2})$ and
obtain the classical equation of motion \r{hs3}. Finally, the
remaining non-vanishing as $\varepsilon\to 0$ terms of order $O(1)$
give us the following equation on $Y^t$:
$$
i\frac{dY^t}{dt}=H_2Y^t,
\l{m4*}
$$
where
$$
H_2= \int D\varphi A^+[\varphi(\cdot)] \int d{\bf x}
\left(-\frac{1}{2}\frac{\delta^2}{\delta \varphi({\bf x})
\delta \varphi({\bf x})}
+\frac{1}{2}(\nabla\varphi_{\Lambda})^2({\bf x})
+ \frac{1}{2}\varphi_{\Lambda}^2({\bf x})
(m^2+(\Phi^t,\varphi_{\Lambda}^2({\bf x})\Phi^t)) \right)
A^-[\varphi(\cdot)]+
$$
$$
+
\frac{\lambda}{4}
\int d{\bf x}
\left[
\int D\varphi \varphi_{\Lambda}^2({\bf x})
(
\Phi^*[\varphi(\cdot)]A^-[\varphi(\cdot)]
+ \Phi[\varphi(\cdot)]A^+[\varphi(\cdot)]
)
\right]^2
\l{m5}
$$

An interesting feature of eq.\r{m5} is that it contains terms with two
creation operators. This means that vacuum vector is not a solution of
eq.\r{m5}. Therefore, the product \r{pr} is not an asymptotic solution
to the $N$-field equation: one cannot suppose the fields to be
independent, it is necessary to take into account the correlatins
between fields even in the leading order of $1/N$-approximation.

\subsubsection{Fixing number of fields}

The constructed asymptotics \r{m2} is a vector of the third-quantized
Fock space. All its components are non-zero. However, our purpose is
to construct the asymptotic solutions to the $N$-field equation. This
means that we should consider the $N$-th component of the vector
\r{m2}. Consider the operator $P_N$ of projecting on the $N$-field
subspace:
$$P_N(\Psi_0,\Psi_1,...,\Psi_N,...)=(0,0,...,0,\Psi_N,0,...).$$
According to subsection 3.2, the vector
$$
P_NU_{\Phi}Y
\l{m6}
$$
is not exponentially small if
$$
\int D\varphi |\Phi[\varphi(\cdot)]|^2 = \varepsilon N.
\l{m7}
$$
The fact that we are interested in the $N$-th component of the vector
\r{m2} only means that expression \r{m2} contains more information
than necessary. This implies that two different vectors $Y_1$ and $Y_2$
may lead to equal $N$-field states $P_NU_{\Phi}Y_1$ and
$P_NU_{\Phi}Y_2$.

Namely, let us use the identity:
$$
P_N \left(\varepsilon \int D\varphi A^+[\varphi(\cdot)]
 A^-[\varphi(\cdot)] - \varepsilon N \right) U_{\Phi}X=0.
\l{m8}
$$
Applying the commutation relations \r{cr2w}, one obtains that
$$
P_N U_{\Phi} \left(
\int D\varphi |\Phi[\varphi(\cdot)]|^2 - \varepsilon N
 + \sqrt{\varepsilon}(a^++a^-)+
 \varepsilon\int D\varphi A^+[\varphi(\cdot)]
 A^-[\varphi(\cdot)] \right) X=0,
$$
where
$$
a^+ =
\int D\varphi A^+[\varphi(\cdot)]
 \Phi[\varphi(\cdot)],
$$
$$
a^- =
\int D\varphi A^-[\varphi(\cdot)]
 \Phi^*[\varphi(\cdot)].
$$
Taking into account eq.\r{m8}, one obtains that for
$$
Y=(a^++a^-)X
$$
the $N$-field state
\r{m6} is small as $\varepsilon\to 0$. This means that there is an
invariance of the vector \r{m2} under transformations
$$
Y \to Y+ (a^++a^-) X.
$$
One can perform the ``gauge-fixing'' procedure by imposing on $Y$ the
additional condition, for example,
$$
a^- \tilde{Y}=0.
\l{m9}
$$
Another approach is to consider the gauge-invariant
generalized state vector
$$
Z=\delta(a^++a^-)Y
\l{m9a}
$$
instead of the Fock space vector $Y$. Eq.\r{m9a} is associated with
the vector $\tilde{Y}$ by the relation
$$
Z=e^{-\frac{1}{2}a^+a^+} \tilde{Y}.
\l{m9*}
$$
Since the operator $a^++a^-$ commutes with the operator
$i\frac{d}{dt}-H_2$, we find that the generalized vector $Z^t$ obeys
the evolution equation:
$$
i\frac{dZ^t}{dt} = H_2 Z^t.
\l{m10}
$$
On the other hand, the vector $\tilde{Y}^t$ does not obey eq.\r{m10}
since  the operator $a^-$ does not commute with
$i\frac{d}{dt}-H_2$. However, the most suitable form of the
$N$-particle wave functional uses the vector $\tilde{Y}$.
Since the vector $U_{\Phi}\tilde{Y}$ can be expressed via components of
$\tilde{Y}$ as follows,
$$
\sum_{n=0}^{\infty} \frac{1}{\sqrt{n!}}
\int \tilde{Y}_n[\varphi_1(\cdot),...,\varphi_n(\cdot)]
A^+[\varphi_1(\cdot)]...A^+[\varphi_n(\cdot)]
D\varphi_1 ... D\varphi_n
\sum_{K=0}^{\infty} \frac{e^{-N/2}}{K!} (a^+)^K |0>,
$$
the expression for the $N$-th component of the asymptotics is:
$$
\Psi_N^t[\varphi_1(\cdot),...,\varphi_n(\cdot)] =
e^{iNS^t} \sum_{n=0}^N \frac{1}{\sqrt{N^nn!}}
\sum_{1\le i_1 \ne ... \ne i_n \le N}
  \tilde{Y}_n[\varphi_{i_1}(\cdot),...,\varphi_{i_n}(\cdot)]
\prod_{i\ne i_1...i_n} \Phi^t[\varphi_i(\cdot)].
\l{m11}
$$
Condition \r{m9} means that the functional $\tilde{Y}_n^t$ is
orthogonal to $\Phi^t$:
$$
\int D\varphi \Phi^{t*}[\varphi_1(\cdot)]
\tilde{Y}_n[\varphi_{i_1}(\cdot),...,\varphi_{i_n}(\cdot)]
=0.
$$
Thus, eq.\r{m11} has the following meaning. The $n=0$-term corresponds
to probability amplitude that all the fields are in the state $\Phi$,
the $n$-th term specifies that amplitude that $N-n$ fields has the
wave functional $\Phi$ etc.

\subsubsection{Renormalization of the cosmological constant}

We have seen in the previous subsection that the phase factor in the
solution to the classical equation $\Phi^t$ diverges. However, the
quantity $S^t$ also diverges. Let us show that these divergences
cancel in a vector $e^{iS^t}\Phi^t$.

First of all, it is necessary to modify the classical equation of
motion, eq.\r{z2}, by adding a cosmological constant $\cal E$ to the
right-hand side:
$$
i\dot{\Phi}_t[\varphi(\cdot)] =
\int d{\bf x}
\left(\frac{1}{2} \pi^2({\bf x}) + \frac{1}{2} (\nabla \varphi_{\Lambda})^2
({\bf x}) + \frac{u_t({\bf x})}{2}\varphi_{\Lambda}^2({\bf x})
+{\cal E}\right)
{\Phi}_t[\varphi(\cdot)].
\l{m12}
$$
This corresponds to usual renormalization of the vacuum energy in the
quantum field theory. Modification \r{m12} means that the $N$-field
Hamiltonian is modified by $N{\cal E}$. Eq.\r{z11} for the singular
pre-exponential factor becomes the following:
$$
\frac{d}{dt}(ln c_t) =
-\frac{1}{2} Tr \hat{R}_t - i\int
d{\bf x} {\cal E}.
\l{m13w}
$$
It follows from eqs.\r{m4} and \r{m13w} that the quantity
$$
a_t = c_t e^{iS_t}
$$
obeys the equation:
$$
\frac{d}{dt} (ln a_t) =
-\frac{1}{2} Tr \hat{R}_t - i\int
d{\bf x} {\cal E}
+\frac{i\lambda}{4} \int d{\bf x} <\varphi_{\Lambda}({\bf x})>^2.
\l{m14w}
$$
Let us find the singularities of $Tr \hat{R}_t$. It is necessary to
expand the symbol of $\hat{R}_t$ up to $O(1/|\omega_{\bf k}|^4)$,
where $\omega_{\bf k}=\sqrt{{\bf k}^2+{\mu}^2}$,
$$
R_t({\bf x},{\bf k}) =
iA_{\bf k} \omega_{\bf k} + r_1 + r_2 +r_3 + O(1/|\omega_{\bf k}|^4),
\l{m15w}
$$
where $r_m=O(1/|\omega_{\bf k}|^m)$.

Substitute expansion \r{m15w} to eq.\r{z10}. Making use of eq.\r{z15a},
we obtain the following relations. The terms of orders $O(1)$,
$O(1/|\omega_{\bf k}|)$ and $O(1/|\omega_{\bf k}|^2)$ are:
$$
2iA_{\bf k}\omega_{\bf k}r_1 + A_{\bf k}*(u-\mu^2)A_{\bf k}=0,
$$
$$
\dot{r}_1 + 2iA_{\bf k} \omega_{\bf k} r_2 -
i\frac{\partial }{\partial \bf k}(iA_{\bf k}\omega_{\bf k})
\frac{\partial r_1}{\partial \bf x}=0,
$$
$$
\dot{r}_2 + 2iA_{\bf k} \omega_{\bf k} r_3
+r_1^2 -
i\frac{\partial }{\partial \bf k}(iA_{\bf k}\omega_{\bf k})
\frac{\partial r_2}{\partial \bf x}
-\frac{1}{2}
\frac{\partial^2}{\partial k_m \partial k_n}
(iA_{\bf k} \omega_{\bf k})
\frac{\partial^2 r_1}{\partial x_m \partial x_n}
=0.
$$
Since trace of the operator is expressed via integral of its symbol
over ${\bf x}$, while integral of the derivative vanish, one finds:
$$
Tr r_2 =  - \frac{d}{dt} Tr \frac{r_1}{2i\omega_{\bf k}},
$$
$$
Tr r_3 = \frac{d^2}{dt^2} Tr \frac{r_1}{4\omega_{\bf k}^2}-
Tr \frac{r_1^2}{2i\omega_{\bf k}}.
$$
We see that the singular part of
$Tr \hat{R}_t$ is presented as:
$$
(Tr \hat{R}_t)_{sing} =
\int d{\bf x} \frac{1}{L^d}\sum_{\bf k}
A_{\bf k}
\left(i\omega_{\bf k}+ \frac{i}{2}\frac{u_t({\bf
x})-\mu^2}{\omega_{\bf k}}
-\frac{i}{8}\frac{(u_t({\bf x})-\mu^2)^2}{\omega_{\bf k}^3}\right)
+\frac{d\cal F}{dt}
$$
for some singular function $\cal F$ depending on $u_t({\bf x})$ and
its derivatives. One can then extract singularities from the factor
$a_t$:
$$
a_t = e^{\cal F} b_t,
$$
where $b_t$ is regular. The fact that the prefactor in the Schrodinger
representation is singular is usual for quantum field theory
\cite{MS7}. However, this singularity vanishes if one changes the
representation \cite{MS7}.

Thus, we see that the cosmological constant should be chosen in order
to make the quantity
$$
-\frac{1}{2L^d} \sum_{\bf k}
A_{\bf k}
\left(i\omega_{\bf k}+ \frac{i}{2}\frac{u_t({\bf
x})-\mu^2}{\omega_{\bf k}}
-\frac{i}{8}\frac{(u_t({\bf x})-\mu^2)^2}{\omega_{\bf k}^3}\right)
+\frac{\lambda}{4}<\varphi_{\Lambda}^2({\bf x})>^2 - {\cal E}
=-\tilde{\cal E}
$$
finite. Making use of the relation \r{z3}
$$
 <\varphi_{\Lambda}^2({\bf x})> =
\frac{u_t({\bf x})-\mu^2}{\lambda}+
\frac{\mu^2-m^2}{\lambda}
$$
and eqs.\r{z23}, \r{z24}, we obtain that the quantity
$-\tilde{\cal E}$ considts of constant and regular parts:
$$
-\tilde{\cal E} =
- {\cal E} + \frac{1}{4\lambda_R}(u-\mu^2)^2
+\frac{1}{4\lambda} (\mu^2-m^2)^2 -
\frac{1}{2L^d} \sum_{\bf k}A_{\bf k}\omega_{\bf k}.
$$
If one sets the bare cosmological constant to be
$$
{\cal E} =
\frac{1}{4\lambda} (\mu^2-m^2)^2 -
\frac{1}{2L^d} \sum_{\bf k}A_{\bf k}\omega_{\bf k},
$$
the divergences in the vacuum energy will be canceled.

\subsection{Non-symmetric solutions to the large-N theory}

The formalism of second quantization works in quantum  mechanics  only
if we  require  the  wave  function  to be symmetric or antisymmetric.
Analogously, the bosonic third quantization procedure is applicable to
the $N$-field   theory   only   if  we  require  the  wave  functional
$\Psi_N[\varphi_1(\cdot),...,\varphi_N(\cdot)]$ to be  symmteric  with
respect to  transpositions  of  the  fields.  This  means  that  third
quantization procedore allows us to construct only symmetric
approximate solutions to the $N$-field Schrodinger equation. One can
notice that the functional \r{m11} is symmetric.

However, one can be interested in more general solutions to the
large-$N$ Schrodinger equation. To construct such solutions, one can
consider the wave functional
$$
\Psi [
\chi_1(\cdot),...,\chi_k(\cdot),\varphi_1(\cdot),...,
\varphi_{N-k}(\cdot)]
$$
to be symmetric with respect to transpositions of the last $N-k$
arguments $\varphi_1$,..., $\varphi_{N-k}$ only.

Let us perform now the procedure of third quantizationonly for
arguments $\varphi$. This means that we consider the extended Fock
space of sets
$$
\left(
\begin{array}{c}
\Psi_0[\chi_1(\cdot),...,\chi_k(\cdot)]\\
\Psi_1[\chi_1(\cdot),...,\chi_k(\cdot),\varphi_1(\cdot)]\\
...\\
\Psi_{N-k}[\chi_1(\cdot),...,\chi_k(\cdot),\varphi_1(\cdot),
...,\varphi_{N-k}(\cdot)])\\
...
\end{array}
\right)
\l{m13}
$$
The $(N-k)$-field Hamiltonian can be presented as a sum of
Hamiltonians of the fields $\varphi$, $\chi$ and interacion term:
$$
H_{N-k}=H_{N-k}^{\varphi} + H^{\varphi\chi} + H^{\chi}_k,
$$
where
$$
H_{N-k}^{\varphi} = \sum_{a=1}^{N-k} \int d{\bf x} \left( -\frac{1}{2}
\frac{\delta^2}{\delta\varphi^a({\bf x})\delta\varphi^a({\bf x})}+
\frac{1}{2}\nabla\varphi_a({\bf x})\nabla\varphi_a({\bf x})+
\frac{m^2}{2}\varphi_a({\bf x})\varphi_a({\bf x})\right)+
$$
$$
+\frac{\lambda\varepsilon}{4} \int d{\bf x}
\sum_{a,b=1}^{N-k} \varphi_a({\bf x})
\varphi_a({\bf x})\varphi_b({\bf x})\varphi_b({\bf x})
$$
is the $\varphi$-Hamiltonian,
$$
H_{N-k}^{\chi} = \sum_{\alpha=1}^{k} \int d{\bf x} \left( -\frac{1}{2}
\frac{\delta^2}{\delta\chi^a({\bf x})\delta\chi^a({\bf x})}+
\frac{1}{2}\nabla\chi_{\alpha}({\bf x})\nabla\chi_{\alpha}({\bf x})+
\frac{m^2}{2}\chi_{\alpha}({\bf x})\chi_{\alpha}({\bf x})\right)+
$$
$$
+\frac{\lambda\varepsilon}{4} \int d{\bf x}
\sum_{\alpha,\beta=1}^{k} \chi_{\alpha}({\bf x})
\chi_{\alpha}({\bf x})\chi_{\beta}({\bf x})\chi_{\beta}({\bf x})
$$
is the Hamiltonian of the $\chi$-field, while the interaction term is
$$
H^{\varphi\chi} = \frac{\lambda\varepsilon}{4}
\int d{\bf x}
\sum_{\alpha=1}^k \chi_{\alpha}({\bf x})\chi_{\alpha}({\bf x})
\sum_{a=1}^{N-k} \varphi_a({\bf x}) \varphi_a({\bf x}).
$$
For the simplicity, the cutoffs are omitted.
The third-quantized version of the Hamiltonian $H_{N-k}^{\varphi}$
was written in subsection 3.1, eq.\r{h2w}. The Hamiltonian $H^{\chi}_k$
is not to be third-quantized. The interaction Hamiltonian is presented
as
$$
H^{\varphi \chi} = \frac{\lambda\varepsilon}{4}
\int d{\bf x}
\sum_{\alpha=1}^k \chi_{\alpha}({\bf x})\chi_{\alpha}({\bf x})
\int D\varphi
A^+[\varphi(\cdot)] \varphi_a({\bf x}) \varphi_a({\bf x})
A^-[\varphi(\cdot)].
$$
One can construct the asymptotics as $N\to\infty$, $k=const$ for the
obtained third-quantized equation by the complex-WKB technique
considered in the previous subsection. The ansatz \r{m2} is
substituted to the Schrodinger equation, where $Y^t$ is a vector
\r{m13}. Equation for $S^t$ and classical equation for $\Phi^t$
coincide with obtained in the previous subsection. Equation for $Y^t$
has the form:
$$
i\frac{dY^t}{dt} = (H_2 + H_{\chi})Y^t,
\l{m14}
$$
where the operator $H_2$ is given by eq.\r{m5}, while the Hamiltonian
$H_{\chi}$ is
$$
H_{\chi} = \sum_{\alpha=1}^{k} \int d{\bf x} \left( -\frac{1}{2}
\frac{\delta^2}{\delta\chi^a({\bf x})\delta\chi^a({\bf x})}+
\frac{1}{2}\nabla\chi_{\alpha}({\bf x})\nabla\chi_{\alpha}({\bf x})+
\frac{m^2
+\lambda (\Phi^t,\varphi^2({\bf x})\Phi^t)
}{2}\chi_{\alpha}({\bf x})\chi_{\alpha}({\bf x})\right).
\l{m15}
$$
We see that $\chi$- and Fock degrees of freedom are splitted in
eq.\r{m14}. Each of fields $\chi$ is a free field interacting with the
external potential $u_t({\bf x})$ \r{z3} specified by the classical
solution $\Phi^t$.

To fix number of fields, one can project the constructed asymptotics
on the $(N-k)$-field subspace. All further derivations are analogous
to the previous subsubsection.

\subsection{Generalization to other models}

We have considered an example of the large-$N$ theory, the
$\Phi^4$-theory. Let us briefly discuss other large-$N$ models and
applications of the third-quantized approach.

\subsubsection{The $\varphi\varphi\chi$-model}

Consider the quantum field theory of $N$ scalar fields
$\varphi^1$,...,$\varphi^N$ and one field $\chi$ with the Lagrangian:
$$
{\cal L}= \frac{1}{2}\partial_{\mu}\varphi^a \partial_{\mu}\varphi^a
-\frac{m^2}{2}\varphi^a\varphi^a + \frac{1}{2}\partial_{\mu}\chi
\partial_{\mu} \chi - \frac{M^2}{2}\chi\chi
-\frac{g}{\sqrt{N}}\varphi^a\varphi^a\chi.
$$
Let us show that semiclassical third-quantized approach is applicable
to this theory and find classical variables and classical equations
for this model.

The Hamiltonian of this model can be expressed via creation and
annihilation operators $A^{\pm}[\varphi(\cdot)]$ in the
third-quantized form:
$$
H=\int D\varphi A^+[\varphi(\cdot)]
\int d{\bf x}
\left[-\frac{1}{2}\frac{\delta^2}
{\delta\varphi ({\bf x}) \delta\varphi ({\bf x})}
+\frac{1}{2}(\nabla\varphi)^2({\bf x}) + \frac{m^2}{2} \varphi^2
({\bf x}) + g\sqrt{\varepsilon} \varphi^2({\bf x}) \chi({\bf x})
\right]
A^-[\varphi(\cdot)] +
$$
$$
+
\int d{\bf x}
\left[-\frac{1}{2}\frac{\delta^2}
{\delta\chi ({\bf x}) \delta\chi ({\bf x})}
+\frac{1}{2}(\nabla\chi)^2({\bf x}) + \frac{M^2}{2} \chi^2
({\bf x})\right].
$$
When one performs the rescaling of ``quantum'' variables,
$$
A^{\pm}\sqrt{\varepsilon} = \Phi^{\pm},
\sqrt{\varepsilon}\chi =Y,
$$
one finds that the Hamiltonian is proportional to $1/\varepsilon$, the
commutator between  operators  $\Phi^{\pm}[\varphi(\cdot)]$  is small,
and the coefficient of each differentiation operator is $\varepsilon$.
This means  that  semiclassical  methods can be applied to this model,
while classical  variables  are  the  following:  complex   functional
$\Phi[\varphi(\cdot)]$, classical  field  $Y({\bf x})$ and canonically
conjugated momentum ${\cal P}({\bf x})$.

The classical  equations  can  be   obtained   by   substituting   the
complex-WKB ansatz
$$
e^{\frac{i}{\varepsilon}S^t} U_{\Phi^t}
\exp\left[\frac{i}{\sqrt{\varepsilon}} \int d{\bf x}
[{\cal P}({\bf x})\chi({\bf x}) - Y({\bf x})
\frac{1}{i} \frac{\delta}{\delta \chi ({\bf x})}
]
\right]Y^t
\l{k1}
$$
to the  Schrodinger equation.  One will obtain the following classical
dynamics:
$$
i\frac{d}{dt}\Phi_t[\varphi(\cdot)]
=
\int d{\bf x}
[-\frac{1}{2}\frac{\delta^2}
{\delta\varphi ({\bf x}) \delta\varphi ({\bf x})}
+\frac{1}{2}(\nabla\varphi)^2({\bf x}) + \frac{m^2}{2} \varphi^2
({\bf x}) + g \varphi^2({\bf x}) Y({\bf x})]
\Phi_t[\varphi(\cdot)],
$$
$$
\dot{Y}={\cal P}, \qquad
-\dot{\cal P} =
-\Delta Y + M^2 Y + g (\Phi, \varphi({\bf x})\Phi).
$$

\subsubsection{Spontaneous symmetry breaking case}

Let us consider the spontaneous symmetry broken large-$N$ field theory,
$$
{\cal L} = \frac{1}{2}\partial_{\mu}\varphi^a
\partial_{\mu}\varphi^a - \frac{\lambda}{4N}
(\varphi^a\varphi^a - Nv^2)^2.
\l{k2}
$$
Since the Hamiltonian can be presented in a third-quantized form,
$$
H=\int D\varphi A^+[\varphi(\cdot)]
\int d{\bf x}
[-\frac{1}{2}\frac{\delta^2}
{\delta\varphi ({\bf x}) \delta\varphi ({\bf x})}
+\frac{1}{2}(\nabla\varphi)^2({\bf x})]
A^-[\varphi(\cdot)] +
$$
$$
+
\frac{\varepsilon\lambda}{4}
\int d{\bf x}
\left(\int D\varphi A^+[\varphi(\cdot)] \varphi^2({\bf x})
A^-[\varphi(\cdot)]- \frac{v^2}{\varepsilon}\right)^2,
$$
our semiclassical approach is applicable to this  case.  However,  the
ansatz \r{m2} seems to correspond to the spontaneous unbroken phase of
the theory  because  fields  $\varphi^1$,  ...,  $\varphi^N$  are  not
distinguished in the asymptotic formula.

To specify  solutions corresponding to the broken phase of the theory,
it is reasonable to denote $\varphi^N=\chi$ and consider the model
$$
{\cal L} =
\frac{1}{2}\partial_{\mu}\chi \partial_{\mu}\chi
+\frac{1}{2}\partial_{\mu}\varphi^a
\partial_{\mu}\varphi^a - \frac{\lambda}{4N}
(\chi^2+\varphi^a\varphi^a - Nv^2)^2.
\l{k3}
$$
One can  then analyse this model analogously to the
previous subsubsection;  the
classical variables will be  the  same:  the  complex  functional
$\Phi[\varphi(\cdot)]$, the field $Y$ and momentum $\cal P$.

\subsubsection{$O(N)$ - nonsymmetric theory}

The method  of  semiclassical third quantization can be applied to the
theories of  a  much  more  wide  types  than  traditional   large-$N$
theories. For example, consider the $O(N)$-nonsymmetric $\Phi^4$-model
$$
{\cal L} =\sum_{a=1}^N
[\frac{1}{2}\partial_{\mu}\varphi^a
\partial_{\mu}\varphi^a -
\frac{m^2}{2}\varphi^a\varphi^a - \frac{g}{4}(\varphi^a)^4]
-
\frac{\lambda}{4N}
(\sum_{a=1}^N\varphi^a\varphi^a )^2.
\l{k4}
$$
Since the third-quantized form of the hamiltonian depends on the small
parameter according  to  eq.\r{m3},  one  can  apply  the  complex-WKB
approach and obtain the following classical equation:
$$
i\dot{\Phi}_t[\varphi(\cdot)] =
\int d{\bf x}
(\frac{1}{2} \pi^2({\bf x}) + \frac{1}{2} (\nabla \varphi)^2
({\bf x}) + \frac{u_t({\bf x})}{2}\varphi^2({\bf x})
+g\varphi^4({\bf x})
)
{\Phi}_t[\varphi(\cdot)],
$$
where
$$
u_t({\bf x})=m^2 + \lambda (\Phi_t,\varphi^2({\bf x})\Phi_t),
(\Phi_t,\Phi_t)=1.
$$
Thus, we see that the classical theory for the large-$N$ system \r{k4}
in the external self-consistent field.  Although  this  model  is  not
exactly solvable,  one  can investigate qualitative features of it and
obtain an information about the model \r{k4}.

\section{Energy spectrum of the large-N theory}

In this section we find stationary asymptotic solutions of the large-N
Schrodinger equation.

\subsection{Approximate stationary solutions}

In the previous section we have constructed the asymptotic solution
\r{m11} to the $N$-field equation. Let us find in what cases
eq.\r{m11} gives us a stationary state:
$$
\Psi_N^t = e^{-iEt} \Psi_N^0.
\l{l0}
$$
First of all, notice that the solution $\Phi^t$ to the classical
equation should depend on $t$ as $e^{-i\Omega t}$:
$$
\Phi^t = e^{-i\Omega t} \Phi^0.
\l{l1}
$$
Then, the $n$-th component of the third-quantized state $\tilde{Y}$ is
also a periodic function:
$$
\tilde{Y}_n^t = e^{-i(n\Omega+\omega)t} \tilde{Y}^0_n.
\l{l2}
$$
If the conditions \r{l1} and \r{l2} are satisfied, eq.\r{m11} gives us
a stationary $N$-field state, because
$$
\frac{dS}{dt} = i(\Phi^t,\dot{\Phi}^t) - H(\Phi^{t*},\Phi^t) =
\Omega - H(\Phi^{0*},\Phi^0),
$$
so that the quantum energy is expressed as:
$$
E=NH(\Phi^*,\Phi)+\omega +O(1/N).
\l{l3}
$$
Let us investigate the obtained conditions, \r{l1} and \r{l2}.
Condition \r{l1} means that one should consider the stationary analog
of eq.\r{z2}:
$$
\int d{\bf x} \left[\frac{1}{2}
\pi^2({\bf x})+ \frac{1}{2}(\nabla\varphi_{\Lambda})^2({\bf x})
+\frac{u({\bf x})}{2}\varphi_{\Lambda}^2({\bf x})\right]
\Phi =\Omega \Phi,
$$
$$
\l{l4}
$$
$$
u({\bf x})=m^2 + \lambda (\Phi, \varphi_{\Lambda}^2({\bf x})\Phi).
$$
To analyze eq.\r{l2}, notice that it can be presented as:
$$
Z_n^t =e^{-i(n\Omega+\omega)t} Z^0_n,
\l{l5}
$$
where the generalized third-quantized vector $Z$ obeying eq.\r{m10} is
related with $\tilde{Y}$ by eq.\r{m9*}. Eq.\r{l5} means that
$$
(H_2Z)_n = (n\Omega+\omega) Z_n.
$$
Making use of the form of the operator of number of fields, one
obtains:
$$
\left(H_2- \Omega \int
D\varphi A^+[\varphi(\cdot)] A^-[\varphi(\cdot)]\right)
Z= \omega Z.
\l{l6}
$$
Eq.\r{m9a} implies the following relation:
$$
\int D\varphi \left(A^+[\varphi(\cdot)]\Phi[\varphi(\cdot)]
+ A^-[\varphi(\cdot)] \Phi^*[\varphi(\cdot)]\right)Z=0.
\l{l7}
$$
Thus, the problem of finding the large-$N$ spectrum is reduced to the
problem of finding a solution to classical equation \r{l4} and
oscillator system \r{l6}, \r{l7} consisting of the oscillator
Schrodinger equation and aonstraint relation \r{l7}.

\subsection{Ground state energy in the large-N theory}

Let us consider the following solution to eq.\r{l4}.  Suppose  $u({\bf
x})$ to be a constant:
$$
u({\bf x})=\mu^2.
$$
Choose the  classical  solution  $\Phi$  to  be  a vacuum state of the
one-field system with mass $\mu$. Thus, our main equation \r{l4} takes
the form:
$$
\mu^2=m^2+\lambda <0|\varphi^2_{\Lambda}({\bf x})|0>.
$$
The firld  $\varphi({\bf  x})$  can  be  expressed  via  creation  and
annihilation operators,
$$
\varphi({\bf x})= \frac{1}{L^{d/2}}\sum_{\bf k}
\frac{\sqrt{A_{\bf k}}}{\sqrt{2\omega_{\bf k}}}
[a_{\bf k}^+e^{-i{\bf k}{\bf x}} +
a_{\bf k}^-e^{i{\bf k}{\bf x}}],
\l{l7*}
$$
where the factor $\sqrt{A_{\bf k}}$ arises because  of  regularization
considered in eq.\r{l4}, while $\omega_{\bf k}=\sqrt{{\bf k}^2+\mu^2}$.
Therefore, one can evaluate the average value of $\varphi^2({\bf x})$:
$$
<0|\varphi^2_{\Lambda}({\bf x})|0>=
\frac{1}{L^d}\sum_{\bf k} \frac{A_{\bf k}}{2\omega_{\bf k}}.
\l{l8}
$$
Thus, we obtain the relation \r{z23} on the parameter $\mu$ playing  a
role of   mass   of   elementary  excitations.  Namely,  non-symmetric
excitations obey Schrodinger equation with  the  Hamiltonian  \r{m15}.
However, this  Hamiltonian  corresponds  to the free field of the mass
$\mu$.

Let us  find  the  leading  order  of  the  energy,  $NH(\Phi^*,\Phi)$
expressed via average values:
$$
H(\Phi^*,\Phi)= <\int d{\bf x}
[
\frac{1}{2}\pi^2({\bf x}) +
\frac{1}{2}(\nabla\varphi_{\Lambda})^2({\bf x})
+\frac{m^2}{2}\varphi_{\Lambda}({\bf x})
]> + \frac{\lambda}{4} <0|\varphi^2_{\Lambda}({\bf x})|0>^2
\l{l8*}
$$
making use of eqs.\r{l4} and \r{l8}, one finds:
$$
E \sim N \left(\Omega - \frac{\lambda}{4L^d}
\left(\sum_{\bf k}
\frac{A_{\bf k}}{2\omega_{\bf k}}\right)^2
\right) +O(1)
\l{l9}
$$
Since the vacuum energy is $\frac{1}{2}\sum_{\bf k}A_{\bf
k}\omega_{\bf k}$, eqs.\r{l9} for the ground-state energy density has
the form:
$$
\frac{E}{NL^d} \to_{N\to\infty}
\frac{1}{2L^d} \sum_{\bf k}A_{\bf k}\omega_{\bf k}
-\lambda
\left(\frac{1}{4L^d}\sum_{\bf k}
\frac{A_{\bf k}}{\omega_{\bf k}}\right)^2
\l{l10}
$$
One can investigate then the dependence of the vacuum energy density
on the boundary conditions.

\subsection{Structure of the ground state wave functional}

Let us find the asymptotic ground state wave functional in the
large-$N$ field theory. It is sufficient to construct a solution to
oscillator-type equation \r{l6} which satisfies the constraint
condition \r{l7}. We will look for this solution in a Gaussian form:
$$
Z=\exp
\left[\frac{1}{2}\int D\varphi D\phi
A^+[\varphi(\cdot)] R[\varphi(\cdot),\phi(\cdot)]
A^+[\phi(\cdot)]
\right]|0>,
\l{l11}
$$
where $|0>$ is a third-quantized vacuum containing no fields. The
constraint \r{l7} leads to the following condition on $R$:
$$
\int R[\varphi(\cdot),\phi(\cdot)] \Phi^*[\phi(\cdot)]
D\phi
= - \Phi[\varphi(\cdot)],
\l{l12}
$$
while Schrodinger equation \r{l6} gives us a nonlinear relation
$$
\sum_{i=1}^2 \left(\int d{\bf x} \left[
-\frac{1}{2}\frac{\delta^2}
{\delta\varphi_i({\bf x}) \delta\varphi_i({\bf x})}
+ \frac{1}{2}(\nabla\varphi_i)^2({\bf x})
+\frac{\mu^2}{2}\varphi_i^2({\bf x})\right] -\Omega\right)
R[\varphi_1(\cdot),\varphi_2(\cdot)]+
$$
$$
+
\frac{\lambda}{2} \int d{\bf x}
\prod_{i=1}^2 \left(
\varphi_i^2({\bf x})\Phi[\varphi_i(\cdot)]
+\int D\tilde{\varphi}_i \tilde{\varphi}_i^2({\bf x})
\Phi^*[ \tilde{\varphi}_i(\cdot)]
R[\varphi_i(\cdot),\tilde{\varphi}_i(\cdot)]\right)=0.
\l{l13}
$$
It follows from eq.\r{m9*} that the third-quantized vector $\tilde{Y}$
entering to eq.\r{m11} has also the Gaussian form:
$$
\tilde{Y} = \exp
\left[\frac{1}{2}\int D\varphi D\phi
A^+[\varphi(\cdot)] R[\varphi(\cdot),\phi(\cdot)]
A^+[\phi(\cdot)]
\right]|0>,
\l{l13*}
$$
where the two-field state $M$ being equal to
$$
M[\varphi(\cdot),\phi(\cdot)]=R[\varphi(\cdot),\phi(\cdot)]
+\Phi[\varphi(\cdot)]\Phi[\phi(\cdot)]
$$
is orthogonal  to  $\Phi[\varphi(\cdot)]$  because of eq.\r{l12}.  The
ground state $N$-field  wave  functional  is  presented  according  to
eq.\r{m11} as
$$
\Psi_N[\varphi_1(\cdot),...,\varphi_N(\cdot)]=
\sum_{l=0}^{[N/2]} \frac{1}{(2N)^ll!}
\sum_{1\le i_1 \ne ...\ne i_{2l} \le N}
M[\varphi_{i_1}(\cdot),\varphi_{i_2}(\cdot)]...
M[\varphi_{i_{2l-1}}(\cdot),\varphi_{i_{2l}}(\cdot)]
\prod_{i\ne i_1...i_{2l} } \Phi[\varphi_i(\cdot)]
\l{l14}
$$
We see  that  one  should take into account the two-field correlations
even in the leading order of the semiclassical approximation.

Let us find the two-field states $M$ and $R$  which  should  obey  the
following condition:   the   norm   of   the   operator   with  kernel
$M[\varphi(\cdot),\phi(\cdot)]$ is  lesser  than  1.  Otherwise,   the
vector \r{l13*} will not belong to the Fock space \cite{Ber}.

It is convenient to consider the following one-field functionals:
$$
\Phi_{{\bf k}_1 {\bf k}_2} =
\frac{1}{\sqrt{2}} a^+_{{\bf k}_1}a^+_{{\bf k}_2}\Phi[\varphi(\cdot)],
\l{l15}
$$
where $\Phi[\varphi(\cdot)]$    is a    vacuum   state   functional
corresponding to  the   field   with   mass   $\mu$.   The   operators
$a^{\pm}_{\bf k}$ are second-quantized creation operators being linear
combinations of field and momentum operators acting in the space of
one-field states.

We are looking for the solution to thenonlinear equation \r{l13} in
the following form:
$$
R[\varphi_1(\cdot),\varphi_2(\cdot)]
=
-\Phi[\varphi_1(\cdot)] \Phi[\varphi_2(\cdot)]
+\sum_{{\bf k}_1{\bf k}_2{\bf p}_1{\bf p}_2}
c_{{\bf k}_1{\bf k}_2{\bf p}_1{\bf p}_2}
\Phi_{{\bf k}_1{\bf k}_2}[\varphi_1(\cdot)]
\Phi_{{\bf p}_1{\bf p}_2}[\varphi_2(\cdot)]
\l{l16}
$$
which automatically obeys the constraint \r{l12}.

Substitution \r{l16} gives us the following equation on the
coefficients
$c_{{\bf k}_1{\bf k}_2{\bf p}_1{\bf p}_2}$:
$$
\sum_{{\bf k}_1{\bf k}_2{\bf p}_1{\bf p}_2}
(A_{{\bf k}_1}\omega_{{\bf k}_1}+
A_{{\bf k}_2}\omega_{{\bf k}_2}+
A_{{\bf p}_1}\omega_{{\bf p}_1}+
A_{{\bf p}_2}\omega_{{\bf p}_2})
c_{{\bf k}_1{\bf k}_2{\bf p}_1{\bf p}_2}
\Phi_{{\bf k}_1{\bf k}_2}[\varphi_1(\cdot)]
\Phi_{{\bf p}_1{\bf p}_2}[\varphi_2(\cdot)]
+
$$
$$
+
\frac{\lambda}{2} \int d{\bf x}
[ (\varphi_{1,\Lambda}^2({\bf x}) - <0|\varphi_{\Lambda}^2({\bf x})|0>)
\Phi[\varphi_1(\cdot)]
+
\sum_{{\bf k}_1'{\bf k}_2'}
c_{{\bf k}_1{\bf k}_2{\bf k}_1'{\bf k}_2'}
\Phi_{{\bf k}_1{\bf k}_2}[\varphi_1(\cdot)]
<0|\varphi^2_{\Lambda}({\bf x})|\Phi_{{\bf k}_1'{\bf k}_2'}>]
$$
$$
\times
[ (\varphi_{2,\Lambda}^2({\bf x}) - <0|\varphi^2_{\Lambda}({\bf x})|0>)
\Phi[\varphi_2(\cdot)]
+
\sum_{{\bf p}_1'{\bf p}_2'}
c_{{\bf p}_1{\bf p}_2{\bf p}_1'{\bf p}_2'}
\Phi_{{\bf p}_1{\bf p}_2}[\varphi_2(\cdot)]
<0|\varphi^2_{\Lambda}({\bf x})|\Phi_{{\bf p}_1'{\bf p}_2'}>]
=0
\l{l17}
$$
Notice that the state vector $\varphi^2_{\Lambda}({\bf x})|0>$
contains only the
vacuum and two-particle components, so that
$$
(\varphi_{2,\Lambda}^2({\bf x}) - <0|\varphi_{\Lambda}^2({\bf x})|0>)
\Phi =
\sum_{{\bf k}_1{\bf k}_2} \Phi_{{\bf k}_1{\bf k}_2}
<\Phi_{{\bf k}_1{\bf k}_2}|\varphi_{\Lambda}^2({\bf x})|0>.
$$
This means that the expression \r{l16} satisfies eq.\r{l13} if
$$
(A_{{\bf k}_1}\omega_{{\bf k}_1}+
A_{{\bf k}_2}\omega_{{\bf k}_2}+
A_{{\bf p}_1}\omega_{{\bf p}_1}+
A_{{\bf p}_2}\omega_{{\bf p}_2})
c_{{\bf k}_1{\bf k}_2{\bf p}_1{\bf p}_2}
+
$$
$$
+
\frac{\lambda}{2}
\int d{\bf x}
\left(<\Phi_{{\bf k}_1{\bf k}_2}|\varphi^2({\bf x})|0>
+
\sum_{{\bf k}_1'{\bf k}_2'}
c_{{\bf k}_1{\bf k}_2{\bf k}_1'{\bf k}_2'}
<0|\varphi^2({\bf x})|\Phi_{{\bf k}_1'{\bf k}_2'}>\right)
$$
$$
\times
\left(<\Phi_{{\bf p}_1{\bf p}_2}|\varphi^2({\bf x})|0>
+
\sum_{{\bf p}_1'{\bf p}_2'}
c_{{\bf p}_1'{\bf p}_2'{\bf p}_1{\bf p}_2}
<0|\varphi^2({\bf x})|\Phi_{{\bf p}_1'{\bf p}_2'}>\right)
=0
\l{l18}
$$
The quantity $c_{{\bf k}_1{\bf k}_2{\bf p}_1{\bf p}_2}$ is a
probability amplitude that there are two particles of one type with
momenta ${\bf k}_1$, ${\bf k}_2$ and of another type with momenta
 ${\bf p}_1$, ${\bf p}_2$. However, the full momentum should be equal
to zero in the ground state. This means that
$c_{{\bf k}_1{\bf k}_2{\bf p}_1{\bf p}_2} \sim
\delta_{{\bf k}_1+{\bf k}_2+{\bf p}_1+{\bf p}_2}$. Let us denote by
${\bf P}$ and $-{\bf P}$ the full momentum of the fields, so that
$$
c_{{\bf k}_1{\bf k}_2{\bf p}_1{\bf p}_2} = \sum_{{\bf P}}
\alpha^{\bf P}_{-{\bf k}_1 \; {\bf p}_1}
\delta_{{\bf k}_1 + {\bf k}_2 +{\bf P}}
\delta_{{\bf p}_1 + {\bf p}_2 -{\bf P}}.
$$
The matrix elements entering to eq.\r{l18} are calculable:
$$
<\Phi_{{\bf k}_1{\bf k}_2}|\varphi_{\Lambda}^2({\bf x})|0>=
\frac{1}{L^d} \frac{
\sqrt{A_{{\bf k}_1}A_{{\bf k}_2}}
}{\sqrt{2\omega_{{\bf k}_1}\omega_{{\bf k}_2}}}
e^{-i({\bf k}_1+{\bf k}_2){\bf x}},
$$
so that eq.\r{l18} is simplified as:
$$
(A_{{\bf k}_1}\omega_{{\bf k}_1}+
A_{{\bf P}-{\bf k}_1}\omega_{{\bf P}-{\bf k}_1}+
A_{{\bf p}_1}\omega_{{\bf p}_1}+
A_{{\bf P}-{\bf p}_1}\omega_{{\bf P}-{\bf p}_1})
\alpha^{\bf P}_{{\bf k}_1{\bf p}_1}
+
$$
$$
+\frac{\lambda}{2L^d}
\sum_{{\bf k}_1'{\bf p}_1'}
\frac{
\sqrt{A_{{\bf p}_1'}A_{{\bf P}-{\bf p}_1'}}
}
{\sqrt{2\omega_{{\bf p}_1'}\omega_{{\bf P}-{\bf p}_1'}}}
\frac{
\sqrt{A_{{\bf k}_1'}A_{{\bf P}-{\bf k}_1'}}
}{\sqrt{2\omega_{{\bf k}_1'}\omega_{{\bf P}-{\bf k}_1'}}}
(\delta_{{\bf k}_1{\bf k}_1'}
+\alpha^{\bf P}_{{\bf k}_1{\bf k}_1'})
(\delta_{{\bf p}_1{\bf p}_1'}
+\alpha^{\bf P}_{{\bf p}_1{\bf p}_1'})=0
\l{l19}
$$

Thus, the equation on the functional $R[\varphi(\cdot),\phi(\cdot)]$
has been reduced to the equation on the function
$\alpha^{\bf P}_{{\bf
k}{\bf p}}$
of two discred variables ${\bf k}$, ${\bf p}$ at fixed
${\bf P}$. To investigate eq.\r{l19}, let us denote by $\alpha^{\bf
P}$ the operator with kernel
$\alpha^{\bf P}_{{\bf k}{\bf p}}$,
by $T^{\bf P}$ we denote the operator of multiplication by
$$
T^{\bf P}=
(A_{{\bf k}}\omega_{{\bf k}}+
A_{{\bf P}-{\bf k}}\omega_{{\bf P}-{\bf k}}),
$$

while $B^{\bf P}$ will be the operator with the following matrix
elemant:
$$
B^{\bf P}_{{\bf k}{\bf p}} =
 \frac{\lambda}{2L^d}
\frac
{
{\sqrt{A_{{\bf k}}A_{{\bf P}-{\bf k}}}}
}
{\sqrt{2\omega_{{\bf k}}\omega_{{\bf P}-{\bf k}}}}
\frac{
{\sqrt{A_{{\bf p}}A_{{\bf P}-{\bf p}}}}
}{\sqrt{2\omega_{{\bf p}}\omega_{{\bf P}-{\bf p}}}}.
$$

Eq.\r{l19} takes the form:
$$
T^{\bf P}\alpha^{\bf P}+\alpha^{\bf P}T^{\bf P}
+(1+\alpha^{\bf P})B^{\bf P}(1+\alpha^{\bf P})=0.
$$

The solution to this operator equation can be written as:
$$
\alpha^{\bf P}=
(1+M^{\bf P})^{-1}(1-M^{\bf P}),
$$
where
$$
M^{\bf P} =
(T^{\bf P})^{-1/2}
[(T^{\bf P})^2 + 2(T^{\bf P})^{1/2}B^{\bf P}
(T^{\bf P})^{1/2}]^{1/2}
(T^{\bf P})^{-1/2}.
\l{l19*}
$$

We see  that the solution to eq.\r{l13} is constructed if the operator
entering to eq.\r{l19*} is positively definite,
$$
\left[(T^{\bf P})^2 + 2(T^{\bf P})^{1/2}B^{\bf P}
(T^{\bf P})^{1/2}\right]^{1/2}>0.
\l{l20}
$$
In this case,  one can construct an approximate stationary solution to
the large-$N$ Schrodinger equation.

Usually, one is able to find semiclassical eigenfunctions  around  the
static classical  solution if it is stable due to small perturbations.
We have seen that investigation of the nonlinear equation gives us  an
alternative way to investigate this problem. We have also obtained the
criteria of stability of classical solution \r{l1}, eq.\r{l20}.

\subsection{Perturbatively excited states}

Let us consider now  the  elementary  excitations  around  the  vacuum
solution. It is necessary to construct excited eigenvectors of
eq.\r{l6}. To construct such a third-quantized vectors, let us find
such an operator $\Lambda^+$ that
$$
[\Lambda^+, H_2 - \Omega \int D\varphi
A^+[\varphi(\cdot)]
A^-[\varphi(\cdot)] ] = -\beta \Lambda^+
\l{l21}
$$
and
$$
[\Lambda^+, \int D\varphi
(
A^+[\varphi(\cdot)]
\Phi[\varphi(\cdot)]
+A^-[\varphi(\cdot)]
\Phi^*[\varphi(\cdot)]
)]=0.
\l{l22}
$$
This operator will satisfy then the following properties. Let $Z$ be a
solution to eq.\r{l6} which satisfies the condition \r{l7}. Eq.\r{l22}
implies that the vector $\Lambda^+Z$ will also obey the constraint
\r{l7}. It follows from eq.\r{l21} that the vector $\Lambda^+Z$ will
be also an eigenstate of eq.\r{l6} with eigenvalue
$$
\omega+\beta.
$$
Thus, the operator $\Lambda^+$ shifts the energy by $\beta$.

Let us find such operators $\Lambda^+$. We are looking for them as
follows:
$$
\Lambda^+(F,G) = i\int D\varphi
(A^+[\varphi(\cdot)]G^*[\varphi(\cdot)]
-A^-[\varphi(\cdot)]F^*[\varphi(\cdot)])
\l{l23}
$$
where $G$ and $F$ are functionals to be found. Since the Hamiltonian
entering to the left-hand side of eq.\r{l6} is quadratic, ehile
operator $\Lambda^+$ linearily depends on creation and annihilation
operators, the commutator will be also a linear combination of these
operators. Thus, eq.\r{l21} will imply the system of two equations:
$$
(\int d{\bf x} [\frac{1}{2}
\pi^2({\bf x})+ \frac{1}{2}(\nabla\varphi_{\Lambda})^2({\bf x})
+\frac{u({\bf x})}{2}\varphi_{\Lambda}^2({\bf x})]
-\Omega)F[\varphi(\cdot)]+
$$
$$
+
\frac{\lambda}{2}\int d{\bf x} \varphi_{\Lambda}^2({\bf x})
\int D\phi \phi_{\Lambda}^2({\bf x}) (F[\phi(\cdot)]
\Phi^*[\phi(\cdot)]+G[\phi(\cdot)]\Phi[\phi(\cdot)])
\Phi[\varphi(\cdot)] = -\beta F[\varphi(\cdot)],
$$
$$
\l{l24}
$$
$$
(\int d{\bf x} [\frac{1}{2}
\pi^2({\bf x})+ \frac{1}{2}(\nabla\varphi_{\Lambda})^2({\bf x})
+\frac{u({\bf x})}{2}\varphi_{\Lambda}^2({\bf x})]
-\Omega)G[\varphi(\cdot)]+
$$
$$
+
\frac{\lambda}{2}\int d{\bf x} \varphi_{\Lambda}^2({\bf x})
\int D\phi \phi_{\Lambda}^2({\bf x}) (F[\phi(\cdot)]
\Phi^*[\phi(\cdot)]+G[\phi(\cdot)]\Phi[\phi(\cdot)])
\Phi^*[\varphi(\cdot)] = \beta G[\varphi(\cdot)],
$$
Thus, we see that spectrum of the variation system \r{l24} coincides
with the spectrum of differences between energy levels, because the
operator $\Lambda^+$ increases the energy by $\beta$.

An important feature is that one muct check the property
$\Lambda^+Z\ne 0$. Otherwise, the operator $\Lambda^+$ will give us a
trivial solution to eq.\r{l6}.

When one applies the operator \r{l23} to the Gaussian state \r{l11},
one will obtain the vector
$$
\int D\varphi A^+[\varphi(\cdot)]
Y[\varphi(\cdot)] Z,
\l{l24*}
$$
where
$$
Y[\varphi(\cdot)]= iG^*[\varphi(\cdot)]
-i \int D\phi R[\varphi(\cdot),\phi(\cdot)]
F^*[\phi(\cdot)].
$$
This functional is certainly non-zero if
$$
(G,G)>(F,F)
\l{l25}
$$
because the norm of the operator $\hat{R}$ with the kernel
$R[\varphi(\cdot),\phi(\cdot)]$ is not greater than 1. Thus, the
solutions of the system \r{l24} which satisfy the condition \r{l25}
will give us the spectrum of elementary excitations.

On the other hand, let us consider the solutions to eq.\r{l24} which
do not obey eq.\r{l25}. Such solutions can be obtained if one
substitute to the system \r{l24} $F^*$ instead of $G$ and $G^*$
instead of $F$. The corresponding operator $\Lambda^+[G^*,F^*]$
denoted as $\Lambda^-[F,G]$ is conjugated to operator \r{l23}.
Consider the vector $\Lambda^-[F,G]Z$. It has the form \r{l24*} where
$$
Y[\varphi(\cdot)]= iF[\varphi(\cdot)]
-i \int D\phi R[\varphi(\cdot),\phi(\cdot)]
G[\phi(\cdot)].
$$
However, one of the definitions of the operator $R$ with the kernel
$R[\varphi,\phi]$ is the following: $RG=F$, see appendix B. This means
that $\Lambda^-[F,G]Z=0$ so that we should take into account only
those solutions to the variation system that obey eq.\r{l25}.

Let us find these solutions. First of all , notice that the functionals
$$
F_{{\bf k}_1...{\bf k}_n}
=0,
G_{{\bf k}_1...{\bf k}_n}
=
a_{{\bf k}_1}^+...a_{{\bf k}_n}^+\Phi^*,
\beta =
A_{{\bf k}_1}
\omega_{{\bf k}_1}+ ... + A_{{\bf k}_n} \omega_{{\bf k}_n}.
\l{l26}
$$
obey eqs.\r{l24} if $n\ne 0, 2$. Namely, the vector
$\varphi^2_{\Lambda}({\bf x})|0>$
contains  vacuum  and two-particle components
only, so that the matrix element
$<F+G|\varphi_{\Lambda}^2({\bf x})|0>$ entering to eq.\r{l24} vanishes.

One can interpret the quantity  $\beta$ as $\Lambda\to\infty$
$$
\omega_{{\bf k}_1}+...
\omega_{{\bf k}_n}
\l{l27}
$$
as an energy of the elementary excitation of the $N$-field system. Why
elementary? One can be surprised that the theory \r{l27} is the energy
of the system of $n$ particles. However, one excitation \r{l26} with
energy \r{l27} is not equivalent to $n$ excitations with energies
$\omega_{{\bf k}_1}$,..., $\omega_{{\bf k}_n}$. Excitation \r{l27}
describes $n$ particles of one type, while $n$ excitations describe
$n$ particles of different types. The third-quantized language is of
course very unusual.

The only non-trivial elementary excitation corresponds to two
particles of one type. Let us look for the corresponding solution to
the variation system as:
$$
G=g_0\Phi^* + \sum_{{\bf k}_1{\bf k}_2}
g_{{\bf k}_1{\bf k}_2}
\Phi^*_{{\bf k}_1{\bf k}_2},
$$
$$
F=f_0\Phi + \sum_{{\bf k}_1{\bf k}_2}
f_{{\bf k}_1{\bf k}_2}
\Phi_{{\bf k}_1{\bf k}_2},
$$
where $\Phi_{{\bf k}_1{\bf k}_2}$ is defined by formula \r{l15}.
Eqs.\r{l24} imply the following nontrivial relations:
$$
(A_{{\bf k}_1}\omega_{{\bf k}_1}+
A_{{\bf k}_2}\omega_{{\bf k}_2}+\beta) f_{{\bf k}_1{\bf k}_2}
+
\frac{\lambda}{2L^d}
\sum_{{\bf p}_1{\bf p}_2}
\frac{
\sqrt{A_{{\bf k}_1}A_{{\bf k}_2}}
}{\sqrt{2\omega_{{\bf k}_1}\omega_{{\bf k}_2}}}
\frac{
\sqrt{A_{{\bf p}_1}A_{{\bf p}_2}}
}
{\sqrt{2\omega_{{\bf p}_1}\omega_{{\bf p}_2}}}
(f_{{\bf p}_1{\bf p}_2}
\delta_{{\bf p}_1+{\bf p}_2,{\bf k}_1+{\bf k}_2 }+
g_{{\bf p}_1{\bf p}_2}
\delta_{{\bf p}_1+{\bf p}_2+{\bf k}_1+{\bf k}_2 }
) =0,
$$
$$
(A_{{\bf k}_1}\omega_{{\bf k}_1}+
A_{{\bf k}_2}\omega_{{\bf k}_2}-\beta) g_{{\bf k}_1{\bf k}_2}
+
\frac{\lambda}{2L^d}
\sum_{{\bf p}_1{\bf p}_2}
\frac{
\sqrt{A_{{\bf k}_1}A_{{\bf k}_2}}
}{\sqrt{2\omega_{{\bf k}_1}\omega_{{\bf k}_2}}}
\frac{
\sqrt{A_{{\bf p}_1}A_{{\bf p}_2}}
}
{\sqrt{2\omega_{{\bf p}_1}\omega_{{\bf p}_2}}}
(g_{{\bf p}_1{\bf p}_2}
\delta_{{\bf p}_1+{\bf p}_2,{\bf k}_1+{\bf k}_2 }+
f_{{\bf p}_1{\bf p}_2}
\delta_{{\bf p}_1+{\bf p}_2+{\bf k}_1+{\bf k}_2 }
) =0.
$$
The next step to simplify the system is to consider the momentum of
two particles to be equal to $\bf P$:
$$
f_{{\bf k}_1{\bf k}_2} = f_{-{\bf k}_1}
\delta_{{\bf k}_1+{\bf k}_2+{\bf P}},
g_{{\bf k}_1{\bf k}_2} = g_{{\bf k}_1}
\delta_{{\bf k}_1+{\bf k}_2-{\bf P}}.
$$
The obtained system
$$
(A_{{\bf k}_1}\omega_{{\bf k}_1}+
A_{{\bf P}-{\bf k}_1}\omega_{{\bf P}-{\bf k}_1}+\beta) f_{{\bf k}_1}
+
\frac{\lambda}{2L^d}
\sum_{{\bf p}_1{\bf p}_2}
\frac{
\sqrt{A_{{\bf k}_1}A_{{\bf P}-{\bf k}_1}}
}
{\sqrt{2\omega_{{\bf k}_1}\omega_{{\bf P}-{\bf k}_1}}}
\frac{
\sqrt{A_{{\bf p}_1}A_{{\bf P}-{\bf p}_1}}
}
{\sqrt{2\omega_{{\bf p}_1}\omega_{{\bf P}-{\bf p}_1}}}
(f_{ {\bf p}_1}
+g_{{\bf p}_1}) =0,
$$
$$
(A_{{\bf k}_1}\omega_{{\bf k}_1}+
A_{{\bf P}-{\bf k}_1}
\omega_{{\bf P}-{\bf k}_1} - \beta) g_{{\bf k}_1}
+
\frac{\lambda}{2L^d}
\sum_{{\bf p}_1{\bf p}_2}
\frac{
\sqrt{A_{{\bf k}_1}A_{{\bf P}-{\bf k}_1}}
}
{\sqrt{2\omega_{{\bf k}_1}\omega_{{\bf P}-{\bf k}_1}}}
\frac{
\sqrt{A_{{\bf p}_1}A_{{\bf P}-{\bf p}_1}}
}
{\sqrt{2\omega_{{\bf p}_1}\omega_{{\bf P}-{\bf p}_1}}}
(f_{{\bf p}_1}
+g_{{\bf p}_1}) =0
$$
allows us to express the functions $f_{\bf k}$ and $g_{\bf k}$ via
eigenvalue $\beta$ and unknown constant $a_{\bf P}$:
$$
f_{\bf k}=-\frac{\lambda}{2L^d}
\frac{a_{\bf P}
\sqrt{A_{\bf k}A_{{\bf P}-{\bf k}}}
}
{\sqrt{2\omega_{\bf k}\omega_{{\bf P}-{\bf k}}}}
\frac{1}{
A_{\bf k}\omega_{\bf k}
+
A_{{\bf P}-{\bf k}}\omega_{{\bf P}-{\bf k}}+\beta}
$$
$$
f_{\bf k}=\frac{\lambda}{2L^d}
\frac{a_{\bf P}
\sqrt{A_{\bf k}A_{{\bf P}-{\bf k}}}
}
{\sqrt{2\omega_{\bf k}\omega_{{\bf P}-{\bf k}}}}
\frac{1}{-
A_{\bf k}\omega_{\bf k}-
A_{{\bf P}-{\bf k}}
\omega_{{\bf P}-{\bf k}}+\beta}
$$
which is determined from the condition
$$
a_{\bf P}= \sum_{\bf p} \frac{f_{\bf p}+g_{\bf p}}
{\sqrt{2\omega_{\bf p}\omega_{{\bf P}-{\bf p}}}}
{\sqrt{A_{\bf p}A_{{\bf P}-{\bf p}}}}
$$
implying equation on $\beta$:
$$
\frac{1}{\lambda} =
\frac{1}{2L^d}
\sum_{\bf p}
\frac{
A_{\bf p} A_{{\bf P}-{\bf p}}
}{\omega_{\bf p} \omega_{{\bf P}-{\bf p}}}
\frac{
A_{\bf p}
\omega_{\bf p}+
A_{{\bf P}-{\bf p}}
\omega_{{\bf P}-{\bf p}}
}
{\beta^2 - (
A_{\bf p}\omega_{\bf p}+
A_{{\bf P}-{\bf p}}
\omega_{{\bf P}-{\bf p}})^2}.
$$
For $d+1=4$, the left-hand and right-hand side of this equation
diverge as $\Lambda\to\infty$.
However, making us of the definition of the renormalized
coupling constant, we obtain the regular relation:
$$
\frac{1}{\lambda_R} =
\frac{1}{2L^d}
(\sum_{\bf p}
\frac{1}{\omega_{\bf p} \omega_{{\bf P}-{\bf p}}}
\frac{\omega_{\bf p}+ \omega_{{\bf P}-{\bf p}}}
{\beta^2 - (\omega_{\bf p}+ \omega_{{\bf P}-{\bf p}})^2}
-\frac{1}{2\omega_{{\bf p}}^3}).
\l{l28}
$$
Eq.\r{l28} allows us to evaluate the energy of bound two-particle
states and scattering amplitudes for these particles.

\subsection{Non-perturbatively excited states}

In the previous subsection we have considered the vacuum state of the
large-$N$ theory and states with finite number of particles. However,
one can also investigate other excited states which cannot be
described by the perturbation theory of the previous subsection.

To understand the problem, consider the simple analogy. In ordinary
quantum field theory at small values of the coupling constant one can
apply the perturbation theory and obtain that excitations correspond
to free particles of given mass in a leading order of perturbation
theory. On the other hand, one can also apply the soliton quantization
approach which is also applicable at small values of coupling constant
and obtain non-perturbative series of states corresponding to the
soliton.

In the large-$N$ theory, the constructed asymptotic energy levels are
found provided that the number of particles is much lesser than $N$.
If we consider the state consisting of $N$ excitations we should take
into account the interaction between the excitations, so that another
approach is necessary.

To consider such a non-perturbatively excited state, it is necessary
to consider another solution to the classical equation \r{l4} rather
than vacuum. For example, the solution
$$
a_{\bf k}^+|0>
$$
($|0>$ is the vacuum of the field system with the mass $M$,
$a_{\bf k}^+$ is a creation operator)
to eq.\r{l4} will correspond to the quantum state of $N$ particles
of a different type with the momentum ${\bf k}$. One will also be able
to find the corresponding Gaussian state, solutions to the variation
system etc.

Let us suppose
$$
u({\bf x})=M^2
$$
and
$$
\Phi = ca_{\bf k_1}^+a_{\bf k_n}^+|0>
\l{w0}
$$
where $c$ is a normalizing factor.
Analogously to section 4.2, one can evaluate the average value of
$\varphi_{\Lambda}^2({\bf x})$:
$$
(\Phi, \varphi_{\Lambda}^2({\bf x})\Phi)=
\frac{1}{L^d} \sum_{\bf k}
\frac{A_{{\bf k}}}{2\sqrt{{\bf k}^2+M^2}}
+\frac{1}{L^d}
\sum_{i=1}^n \frac{A_{{\bf k}_i}}{\sqrt{{\bf k}_i^2+ M^2}}.
\l{w1}
$$
Eq. \r{l4} on the mass $M$
is to be renormalized. Making use of the definition of the physical
mass $\mu$ of perturbative excitations, one obtains:
$$
\frac{M^2-\mu^2}{\lambda}=
\frac{1}{2L^d} \sum_{\bf k} A_{\bf k}
(\frac{1}{\sqrt{{\bf k}^2+M^2}}
-\frac{1}{\sqrt{{\bf k}^2+{\mu}^2}})+
\frac{1}{L^d}\sum_{i=1}^n \frac{
A_{{\bf k}_i}}{\sqrt{{\bf k}^2_i+M^2}}
$$
For $d=1,2$ the right-hand and left-hand sides of this equation are
finite. For d=3, it is necessary to perform the renormalization of the
coupling constant:
$$
\frac{M^2-\mu^2}{\lambda_R}=
\frac{1}{2L^d} \sum_{\bf k} A_{\bf k}
(\frac{1}{\sqrt{{\bf k}^2+M^2}}
-\frac{1}{\sqrt{{\bf k}^2+{\mu}^2}}
+\frac{M^2-\mu^2}{2({\bf k}^2+\mu^2)^{3/2}}
)+
\frac{1
}{L^d}\sum_{i=1}^n \frac{
A_{{\bf k}_i}}{\sqrt{{\bf k}^2_i+M^2}}
$$
The quantity $M$ plays a role of a mass of small excitations around
the background state corresponding to the classical functional \r{w0}.
It differs from $\mu$.

To find the leading order of the energy of the quantum state, let us
use eq.\r{l8*}. One obtains:
$$
\frac{E}{N}=\Omega - \frac{\lambda}{4} \int d{\bf x}
<\varphi^2_{\Lambda}({\bf x})>^2,
$$
so that
$$
\frac{E}{N}=\frac{1}{2}\sum_{{\bf k}} A_{\bf k}
\sqrt{{\bf k}^2+M^2} + \sum_{i=1}^n
A_{{\bf k}_i}\sqrt{{\bf k}^2_i+M^2}
-\frac{\lambda L^d}{4} \left(\frac{M^2-m^2}{\lambda}\right)^2,
$$
where we have used eq.\r{l4}. The difference between this energy and
ground state energy \r{l10} is
$$
\frac{E-E_0}{N} =
\frac{1}{2}
\sum_{{\bf k}} A_{\bf k}
(\sqrt{{\bf k}^2+M^2}-
\sqrt{{\bf k}^2+\mu^2})
+ \sum_{i=1}^n A_{{\bf k}_i}\sqrt{{\bf k}^2_i+M^2}
-\frac{\lambda L^d}{4\lambda} (M^2-\mu^2)
(M^2+\mu^2-2m^2),
$$
since
$$
M^2+\mu^2-2m^2=
\frac{\lambda}{L^d}
(\sum_{\bf k} \frac{A_{\bf k}}{2}
(\frac{1}{\sqrt{{\bf k}^2+M^2}}
+\frac{1}{\sqrt{{\bf k}^2+\mu^2}})
+ \sum_{i=1}^n
\frac{A_{{\bf k}_i}}{\sqrt{{\bf k}^2_i+M^2}}
)
$$
The energy of non-perturbative excitation is
$$
\frac{E-E_0}{N} =
\frac{1}{2}
\sum_{{\bf k}} A_{\bf k}
(\sqrt{{\bf k}^2+M^2}-
\sqrt{{\bf k}^2+\mu^2}
-\frac{M^2-\mu^2}{4}
(\frac{1}{\sqrt{{\bf k}^2+M^2}}
+\frac{1}{\sqrt{{\bf k}^2+\mu^2}})
)
$$
$$
+ \sum_{i=1}^nA_{{\bf k}_i} (\sqrt{{\bf k}^2_i+M^2}-
\frac{M^2-\mu^2}{4}
\frac{1}{\sqrt{{\bf k}^2_i+M^2}})
\l{w2}
$$
It is interesting that for $d\ge 5$ the sum entering to eq.\r{w2} is
divergent. This corresponds to the fact that the $\phi^4$-theory in
such dimensions is not renirmalizable. Usually this feature of the
theory is important in calculations of the higher orders of
perturbation theory. We see that for the method of the
large-$N$ expansion the feature of the renormalizability
is very important even in the leading order.

One can also consider small excitations around the considered
non-perturbative state. One should investigate then the variation
system which has the form \r{l24} where $\Phi$ is not the vacuum state
functional but the functional \r{w0}. There will be trivial solutions
to eq.\r{l24} with $F=0$, this is the case
$$
<G^*|\varphi_{\Lambda}^2({\bf x})|0> =0.
$$
In this case the eigenvalue $\beta$ will coincide with one of the
eigenvalues of the operator $H_M-\Omega$ ($H_M$ is the Hamiltonian of
the field of mass $M$).

To find non-trivial values of $\beta$, denote by $F$ and $X$ the
one-field states corresponding to the functionals $F[\varphi(\cdot)]$
and $G^*[\varphi(\cdot)]$. Eqs.\r{l24} will take then the form:
$$
(H_M-\Omega)F + \frac{\lambda}{2} \int d{\bf x}
\varphi_{\Lambda}^2({\bf x}) (<X|\varphi_{\Lambda}^2({\bf x})|\Phi>
+ <\Phi|\varphi_{\Lambda}^2({\bf x})|F>) \Phi =-
\beta F
$$
$$
\l{w3}
$$
$$
(H_M-\Omega)X + \frac{\lambda}{2} \int d{\bf x}
\varphi_{\Lambda}^2({\bf x}) (<\Phi|\varphi_{\Lambda}^2({\bf x})|X>
+ <F|\varphi_{\Lambda}^2({\bf x})|\Phi>) \Phi =
\beta X
$$
Let ${\bf K}={\bf k}_1+...+{\bf k}_n$. To simplify the system \r{w3},
consider the stste $X$ to have momentum ${\bf P}$ and $F$ to have
momentum $2{\bf K}-{\bf P}$: for momentum operator $\hat{\bf P}=
\sum_{\bf k} {\bf k}a_{\bf k}^+ a_{\bf k}^-$  one has:
$$
\hat{\bf P}X={\bf P}X,\quad \hat{\bf P}F=(2{\bf K}-{\bf P})F.
$$
This condition implies that
$$
<X|\varphi_{\Lambda}^2({\bf x})|\Phi> =
<X|e^{-i\hat{\bf P}{\bf x}}\varphi_{\Lambda}^2(0)
e^{i\hat{\bf P}{\bf x}}|\Phi> =
e^{i({\bf K}-{\bf P}){\bf x}}
<X|\varphi_{\Lambda}^2(0)|\Phi>,
$$
while
$$
<\Phi|\varphi_{\Lambda}^2({\bf x})|F> =
e^{i({\bf K}-{\bf P}){\bf x}}
<\Phi|\varphi_{\Lambda}^2(0)|F>.
$$
Analogously, one can express the operator
$\int d{\bf x}e^{i{\bf k}{\bf x}}\varphi^2({\bf x})$ via the
projection operators $\Pi_{\bf Q}$ on the subspace corresponding to a
given momentum $\bf Q$,
$$
\int d{\bf x} e^{i({\bf K}-{\bf P}){\bf x}}
\varphi_{\Lambda}^2({\bf x})\Phi=
L^d \Pi_{2{\bf K}-{\bf P}} \varphi^2(0) \Phi.
$$
Thus, one simplifies the variation system as
$$
(H_M-\Omega+\beta)F + \frac{\lambda L^d a^*}{2}
\Pi_{2{\bf K}-{\bf P}} \varphi_{\Lambda}^2(0)\Phi=0,
$$
$$
(H_M-\Omega-\beta)X + \frac{\lambda L^d a}{2}
\Pi_{{\bf P}} \varphi_{\Lambda}^2(0)\Phi=0,
$$
$$
a=<\Phi|\varphi_{\Lambda}^2(0)|X> + <F|\varphi_{\Lambda}^2(0)|\Phi>,
$$
so that the nontrivial eigenvalues $\beta$ are expressed from the
equation
$$
\frac{1}{\lambda} =
\frac{1}{L^d}
<\Phi| \varphi_{\Lambda}^2(0)
[(\beta-H_M+\Omega)^{-1}\Pi_{\bf P}-
(\beta+H_M-\Omega)^{-1}\Pi_{2{\bf K}-{\bf P}}]
\varphi_{\Lambda}^2(0)|\Phi>
$$
analogously to the previous subsection.

\section{Conclusions}

The method  of  second  quantization  is  very   useful   in   quantum
many-particle mechanics even in the case of a fixed number of
particles. For example, this approach allows us to introduce a notion
of quasiparticles which can be created or annihihated even if all the
particles of the system are stable. The quasiparticle conception is
very important for the condensed matter theory.

Analogously, we have seen that the notion of third quantization can be
applied to the theory of $N$ fields and allows us to construct the
approximate solutions to the $N$-field functional Schrodinger
equation. These asymptotics are expressed via the solution to eqs.
\r{hs3} and \r{m5}.

Although eq. \r{hs3} is a classical equation, it resembles a
functional Schrodinger equation of quantum theory of one field rather
than a classical field equation. Eq. \r{m5} is an equation for the
vector of the third-quantized Fock space vector. The corresponding
Hamiltonian is expressed via the operators $A^{\pm}[\varphi(\cdot)]$
which can be called as operators of creation and annihilation of the
``quasifield'' since the analogous operators in quantum statistics
create and annihilate quasiparticles. The quantum theory of fixed
number of fields is reduced to the theory of variable number of fields
\r{m5}. Since the Hamiltonian is quadratic, such a model is
exactly-solvable by the Bogoliubov-like transformation. However, the
coefficients of the transformation are functionals expressing via the
variation system \r{l24}.

In quantum field theory it is also assumed that the set of elementary
particles is fixed: there are electrons, muons etc. There were no
attempts to construct a theory admitting existence of the electron
field with some probability. However, we see that the theory \r{m5}
lells us that there are probability amplitudes that there are no
fields, that there is only one type of particles, etc. Since the
theory \r{m5} has been shown to be equivalent to the ordinary
$N$-field theory, the model \r{m5} obeys all the properties of the
field theory, for example, it is relativistic invariant etc.

The conception that particles can be created or annihilated was very
important in constructing the quantum field theory. One can hope that
the idea that number of fields can be variable will be also useful in
developing further theories of everything.

This work was supported by the Russian Foundation for Basic Research,
project 96-01-01544.

\section*{Appendix A. The complex-WKB semiclassical approach}

In this appendix we briefly review the complex-WKB method
which was developed in \cite{M2}.

\subsection*{A.1. The WKB and complex-WKB ansatz}

Semiclassical approximation  is  the  powerful   tool   to   construct
asymptotic solutions   to   the   quantum  mechanical  $d$-dimensional
equations like
$$
i\hbar \frac{\partial \psi}{\partial t} =
H\left(x, -i\hbar \frac{\partial}{\partial x}\right)\psi
\l{s5}
$$
for the wave function $\psi^t(x)$ as $\hbar\to 0$.  We have seen  that
equations like  \r{s5}  arise  in the large-N field theory,  while the
small parameter $\hbar$ may be not related with  the  Planck  constant
$\hbar$: in the large-$N$ theory $\hbar=1/N$.

The most  famous  semiclassical  approach  is  the  WKB-approach which
allows us to construct semiclassical solutions  to  eq.\r{s5}
of the rapidly oscillating form
$$
\varphi^t(x) e^{\frac{i}{\hbar}S^t(x)}
\l{sol1}
$$
where $S$ is a real function.  One can  easily  see  that  the  ansatz
\r{sol1} really satisfies
eq.\r{s5}, obtain the Hamilton-Jacobi equation for $S$,
the expansion for $\varphi$ etc.

However, there  exists  other  semiclassical  wave functions that also
approximately obey eq.\r{s5} as $\hbar\to 0$.  One more example of the
semiclassical ansatz to eq.\r{s5} is the complex-WKB ansatz \cite{M2}:
$$
\psi^t(x)=\frac{1}{\hbar^{d/4}}e^{\frac{i}{\hbar}                 S^t}
e^{\frac{i}{\hbar}P^t(x-Q^t)}f^t(\frac{x-Q^t}{\sqrt{\hbar}})
\l{sol2}
$$
where $f^t(\xi)$ is a rapidly damping function of $\xi$,
while $S^t$,  $P^t$,  $Q^t$ do not depend on $x$,  $P^t$ and $Q^t$ are
$d$-dimensional vectors.
The normalizing factor $\hbar^{-d/4}$ is a corollary of the  condition
$(\psi,\psi)=O(1)$.

At fixed moment of time the
wave  function  \r{sol2}  is a wave packet with the width of order
$O(\sqrt{\hbar})$.  The  average  value  of  the coordinate is
$Q^t$, average momentum is $P^t$,  uncertainties of the coordinate and
momentum are of order $\sqrt{\hbar}$,  so that the uncertainty relation
$\delta P \delta Q \sim \hbar$ is satisfied.  Note that  the  WKB-wave
function has  the  uncertainty of the coordinate and momentum of order
$O(1)$. Thus,  the wave function \r{sol2} really determines motion  of
the classical  particle  along  the  classical  trajectory,  while the
WKB-function \r{sol1} does not determine a classical trajectory.

Let us show that  the  wave  function  \r{sol2}  really  approximately
satisfies the  evolution  equation  \r{s5}  First of all,  notice that
extraction of the multiplier
$$
e^{\frac{i}{\hbar}S^t}e^{\frac{i}{\hbar}P^t(x-Q^t)}
$$
from the wave function $\psi^t$,
$$
\psi^t(x)=
e^{\frac{i}{\hbar}S^t}e^{\frac{i}{\hbar}P^t(x-Q^t)}
\chi^t(x)
$$
is equivalent to shifting the differential operators:
$$
-i\hbar\frac{\partial}{\partial x}
e^{\frac{i}{\hbar}S^t}e^{\frac{i}{\hbar}P^t(x-Q^t)}
=
e^{\frac{i}{\hbar}S^t}e^{\frac{i}{\hbar}P^t(x-Q^t)}
(P^t-i\hbar\frac{\partial}{\partial x}),
$$
$$
\l{com}
$$
$$
i\hbar\frac{\partial}{\partial t}
e^{\frac{i}{\hbar}S^t}e^{\frac{i}{\hbar}P^t(x-Q^t)}
=
e^{\frac{i}{\hbar}S^t}e^{\frac{i}{\hbar}P^t(x-Q^t)}
(i\hbar\frac{\partial}{\partial t}-\dot{S}^t-\dot{P}^t(x-Q^t)
+P^t\dot{Q}^t).
$$
Identities \r{com}  for operators can be justified by applying them to
arbitrary test function. Eqs.\r{com} imply that
$$
(i\hbar\frac{\partial}{\partial t}-\dot{S}^t-\dot{P}^t(x-Q^t)
+P^t\dot{Q}^t - H(x,P^t-i\hbar\frac{\partial}{\partial x}))
\chi^t(x)=0.
\l{chi}
$$
Let us  consider  the  limit  $\hbar\to 0$ of the latter equation.  At
first sight,  one should  simply  neglect  all  the  terms  containing
$\hbar$, so  that  one  would  wonder  why the function $S^t$ does not
depend on $x$.  However,  one should take into account that  the  wave
function $\chi^t(x)$   is   a   wave   packet   with  width  of  order
$\sqrt{\hbar}$, so that
$$
x \chi^t \simeq Q^t \chi^t.
$$
This means that one should not only set  $\hbar=0$  in  eq.\r{chi}  in
order to obtain a leading order in $\hbar$ but  also  set  $x=Q^t$  in
this equation,  so  that the leading semiclassical approximation gives
us the following relation on $S^t$:
$$
\dot{S}^t=P^t\dot{Q}^t - H(Q^t,P^t),
\l{act1}
$$
so that  $S^t$  is  the action on the trajectory of motion of the wave
packet. To find next corrections, consider the substitution
$$
x-Q^t=\xi\sqrt{\hbar},
$$
so that the average number of the observable $\xi$ is of order $O(1)$.
Using eq.\r{act1}, one finds that eq.\r{chi} takes the form
$$
(i\hbar\frac{\partial}{\partial t}-\dot{P}\xi\sqrt{\hbar}
-i\sqrt{\hbar}\frac{\partial}{\partial \xi}\dot{Q}^t
-H(Q^t+\sqrt{\hbar}\xi, P^t-i\hbar\frac{\partial}{\partial \xi}))
f^t(\xi)=0.
\l{f1}
$$
Eq. \r{f1} imply that $(P^t,Q^t)$ should obey the Hamiltonian system:
$$
\dot{Q}^t=\frac{\partial H}{\partial P},
\dot{P}^t=-\frac{\partial H}{\partial Q},
\l{ham}
$$
while the  dynamical  equation for $f^t$ is the oscillator Schrodinger
equation:
$$
(i\frac{\partial}{\partial t}-H_2)f^t=0,
\l{f2}
$$
where
$$
H_2=
\frac{1}{2}\xi\frac{\partial^2H}{\partial Q \partial Q}\xi
+
\frac{1}{2}\xi\frac{\partial^2H}{\partial Q \partial P}
\frac{1}{i}\frac{\partial}{\partial \xi}
+
\frac{1}{2}
\frac{1}{i}\frac{\partial}{\partial \xi}
\frac{\partial^2H}{\partial P \partial P}
\frac{1}{i}\frac{\partial}{\partial \xi}
\l{h2}
$$
The arguments $(Q^t,P^t)$ of the classical Hamiltonian are omitted.

Not that  for $\hbar$-dependent Hamiltonian,  $H(Q,P)+\hbar H_1(Q,P)$,
the classical equations \r{ham} will not change,  while the oscillator
Hamiltonian \r{h2} will contain the additional term, $H_1(Q^t,P^t)$.

We see  that  the complex-WKB approach is a good method to justify the
Ehrenfest theorem.  Moreover,  one can find  not  only  the  classical
trajectory of  motion of the wave packet but also the evolution of the
shape of this wave packet.  Let us show how the  complex-WKB  approach
allows us to obtain the asymptotic spectrum of the Hamiltonian.

\subsection*{A.2. Stationary complex-WKB solutions}

Let us  find  ``semiclassical'' stationary states with the help of the
complex-WKB approach.  Since the  wave  function  \r{sol2}  should  be
stationary, the  wave  packet  should  not  move,  so  that  classical
coordinate and  momentum  should  be  time-independent,   $P^t=const$,
$Q^t=const$. The phase factor depends on $t$ as
$$
S^t=const - H(Q,P) t.
$$
The wave      function     \r{sol2}{     depends     on     $t$     as
$e^{-\frac{i}{\hbar}{\cal E} t}$ for
$$
{\cal E}= H(Q,P) + \hbar\varepsilon,
$$
if
$$
H_2 f = \varepsilon f.
\l{s7}
$$
We see  that  the  problem  of finding approximate energy levels of an
arbitrary Hamiltonian is reduced  to  the  problem  of  finding  exact
spextrum of   the  oscillator Hamiltonian  $H_2$.  We  see  that  the
oscillator approximation  considered  in   \cite{sol}   is   really
semiclassical.

The procedure  of  the  diagonalization  of  the quadratic Hamiltonian
\r{h2} is standard \cite{Ber}.  One should  consider  $d$  independent
creation and annihilation operators
$$
A_k^-=p^{(k)}\xi - q^{(k)} \frac{1}{i} \frac{\partial}{\partial \xi},
$$
$$
\l{ac}
$$
$$
A_k^+=p^{(k)*}\xi - q^{(k)*} \frac{1}{i} \frac{\partial}{\partial \xi},
$$
obeying the canonical commutation relations
$$
[A_k^-,A_l^+]=\delta_{kl}, [A_k^-,A_l^-]=0.
$$
which are equivalent to
$$
p^{(k)}q^{(l)*} -
q^{(k)}p^{(l)*} =i\delta_{kl},
$$
$$
p^{(k)}q^{(l)} -
q^{(k)}p^{(l)} =0.
$$
The $d$-dimensional vectors $p^{(k)}$ and $q^{(k)}$ entering to
creation and annihilation operators \r{ac} are defined from the
relations
$$
[H_2, A_k^{\pm}] = \pm \omega_k A_k^{\pm}.
\l{cr2}
$$
Since there are $2d$ linearily independent operators $A^{\pm}$, one
can present $2d$ coordinate and momenta operators via creation
andannihilation operators. Thus, eq.\r{cr2} means that the operator
$H_2-\sum_k \omega_k A_k^+A_k^-$ commutes with coordinate and momentum
operators. therefore, it is a c-number,so that
$$
H_2=\sum_k \omega_k A_k^+A_k^-+\varepsilon^{(0)}.
$$
The ``frequaences'' $\omega_k$ are obtained from eqs.\r{cr2} being
equivalent to the variation system to the classical equations:
$$
-i\omega p = \frac{\partial^2 H}{\partial Q \partial P} p +
 \frac{\partial^2 H}{\partial Q \partial Q} q,
$$
$$
i\omega q = \frac{\partial^2 H}{\partial P \partial P} p +
 \frac{\partial^2 H}{\partial P \partial Q} q,
$$
One can show \cite{M2} that the descibed procedure of diagonalization
is well-defined if and only if the stationary solution $(P,Q)$ of the
classical equations of motion is stable.

The wave function of the ground state is defined from the condition
$$
A_1^- f =0,..., A_d^- f=0.
$$
The function $f$ has the Gaussian form
$$
f=\exp(\frac{i}{2}\xi\alpha\xi),
$$
where
$$
p^{(k)}=\alpha q^{(k)}
\l{bc}
$$
one can show that the matrixconsisting of $d$ vectors $q^{(1)}$,
...,$q^{(d)}$ is reversible, so that eq. \r{bc} determines the matrix
$\alpha$ in a unique fashion. The ground state energy can be written as
$$
\varepsilon^{(0)}= \frac{1}{2i} Tr (
\frac{\partial^2 H}{\partial Q \partial P}\alpha)
+H_1.
\l{e0}
$$
The excited states are found by applying creation operators to the
ground state
$$
(A_1^+)^{n_1}...(A_d^+)^{n_d} f_0.
\l{b19}
$$
and have energies
$$
\varepsilon^{(0)} + \sum \omega_kn_k.
\l{b20}
$$
We have shown that the stationary case of the complex-WKB approach
leads to the oscillator approximation.

\section*{Appendix B. The complex-WKB method in Fock space}

In this appendix we develop the complex-WKB method for the
second-quantized equations. For more details, see \cite{MS1}.

\subsection*{B.1 The complex-WKB ansatz}

Consider the Schrodinger equation
$$
i\frac{d\Psi^t}{dt}=\hat{H}\Psi^t
\l{b1}
$$
for the time-dependent vector $\Psi$ of the Fock space. It happens
that semiclassical methods are applicable to eq.\r{b1} if the
Hamiltonian depends on the bosonic creation and annihilation operators
and small parameter $\varepsilon$ as follows:
$$
\hat{H} = \frac{1}{\varepsilon}
H(\sqrt{\varepsilon}a^+,\sqrt{\varepsilon}a^-)
\l{b2}
$$
For the case of a finite number of degrees of freedom one can consider
the following representation for the creations and annihilation
operators. The vacuum state corresponds to the wave function
$$
\Psi_0(\xi)=\exp(-\frac{1}{2\varepsilon}\sum_i x_i^2),
\l{b2*}
$$
while the operators $a^{\pm}$ can be presented as
$$
a_i^{\pm}= \frac{x_i \mp \varepsilon \partial/\partial x_i}
{\sqrt{2\varepsilon}}.
\l{b3}
$$
Substituting eq.\r{b3} to eq.\r{b2}, one finds that the Schrodinger
equation \r{b1} is taken to the semiclassical form:
$$
i\varepsilon\frac{\partial\psi}{\partial t} =
H(
\frac{x-\partial/\partial x}{\sqrt{2}},
\frac{x+\partial/\partial x}{\sqrt{2}}
)
\psi,
$$
while the analog of the Planck constant is $\varepsilon$,
$\varepsilon=\hbar$.

We see that one can apply the complex-WKB method to the Hamiltonian of
the type \r{b2}. Note that $N$-field Hamiltonians considered in this
paper depend on the small parameter analogously to \r{b2}.

To construct an analog of the wave packet \r{sol2}, let us consider
the partial case of the formula \r{sol2}, the particle eith zero
coordinate and momentum, with the wave function
$$
f(x/\sqrt{\varepsilon}).
$$
It is easy to see that this quantum state corresponds to an
$\varepsilon$-independent state vector of the Fock space. Really, such
a vector can be expressed via creation operators and vacuum state
\r{b2*}. However, the wave function \r{b2*} depends on
$x/\sqrt{\varepsilon}$ only, while the creation operators are
expressed via operators of multiplication by $x/\sqrt{\varepsilon}$
and operator of differentiation with respect to
$x_i/\sqrt{\varepsilon}$. This means that $\varepsilon$-independent
state describes a wave packet with $P=0$, $Q=0$.

To construct the Fock-space analog of the wave function \r{sol2} in
general case, present it as
$$
const
e^{\frac{i}{\varepsilon}(Px-Q\frac{1}{i}\frac{\partial}{\partial x})}
f(\frac{x}{\sqrt{\varepsilon}}).
$$
Since operators $x$ and $\partial/\partial x$ are linear combinations
of creation and annihilation operators, the exponent can be presented
as
$$
\exp(\frac{1}{\sqrt{\varepsilon}}\sum_k
(
\varphi_k a^+_k - \varphi_k^*a^-_k
))
$$
for some numbers $\varphi_i$. Thus, we see that the complex-WKB ansatz
in the Fock space should depend on $\varepsilon$ as follows:
$$
\Psi^t =
U_{\varphi^t,S^t}
Y^t,
\l{b4}
$$
where
$$
U_{\varphi,S}
=
e^{\frac{i}{\varepsilon}S}
e^{\frac{1}{\sqrt{\varepsilon}}\sum_k
(\varphi_k a^+_k - \varphi_k^*a^-_k )}.
$$
This ansatz can be considered also in the infinite-dimensional case.

Analogously to appendix A, one can show that vector \r{b4}
approximately satisfies eq.\r{b1}. Namely, the following relations
take place:
$$
a^+_k U_{\varphi,s} = U_{\varphi,s} (a^+_k +
\frac{\varphi^*_k}{\sqrt{\varepsilon}})
$$
$$
\l{b4*}
$$
$$
a^-_k U_{\varphi,s} = U_{\varphi,s} (a^-_k +
\frac{\varphi_k}{\sqrt{\varepsilon}})
$$
$$
\l{b5}
$$
$$
U^{-1}_{\varphi,s} i\frac{d}{dt}
U_{\varphi,s} =
(-\frac{1}{\varepsilon}\frac{dS^t}{dt}
+\frac{i}{2\varepsilon}
\sum_k(\varphi_k^* \frac{d\varphi_k}{dt}
- \frac{d\varphi_k^*}{dt} \varphi_k ) +
\frac{i}{\sqrt{\varepsilon}} \sum_k
(\dot{\varphi}_k a_k^+ - \dot{\varphi}_k^*a_k^-)
$$
In the leading order $O(1/\varepsilon)$ we obtain the condition on
$S^t$
$$
\dot{S}=\frac{i}{2}
\sum_k(\varphi^*_k \frac{d\varphi_k}{dt}
- \frac{d\varphi_k^*}{dt} \varphi_k ) -
H(\varphi^*,\varphi).
\l{b5*}
$$
The next order $O(1/\sqrt{\varepsilon})$ leads us to classical
equations for classical variables $\varphi_k^*$:
$$
i\frac{d\varphi_k}{dt}=
\frac{\partial H}{\partial \varphi_k^*},
\l{b6}
$$
while the terms of order $O(1)$ allow us to obtain the equation for
the Fock space vector $Y^t$:
$$
i\frac{dY^t}{dt} =
H_2Y^t.
\l{b6*}
$$
where
$$
H_2=\sum_{k,l}
[
\frac{1}{2}
a_k^+ \frac{\partial^2H}{\partial \varphi^*_k\partial
\varphi^*_l} a^+_l
+
a_k^+ \frac{\partial^2H}{\partial \varphi^*_k\partial
\varphi_l} a^-_l
+
\frac{1}{2}
a_k^- \frac{\partial^2H}{\partial \varphi_k\partial
\varphi_l} a^-_l
]
\l{b6+}
$$

\subsection*{B.2 Asymptotic spectrum of the $N$-particle Hamiltonian}

Let us consider the Hamiltonians \r{b2} that consists of terms with
equal number of creation and annihilation operators, i.e. the
Hamiltonians that are invariant under the global transformations,
$a^{\pm}_{\bf k}\to a^{\pm}_{\bf k}e^{\pm i\alpha}$. This means that
the numbar of particles is an integarl of motion.

To find eigenstates of the $N$-particle component of the Hamiltonian,
consider the $N$-th component of eq.\r{b4} and find find whether it is
a stationary   vector.  For  the  simplicity,  set  $N=1/\varepsilon$.
Eqs.\r{b4*} implies  that  the  average  value   of   any   observable
$f(\varepsilon \sum_m     a^+_ma^-_m)$,     where     $f$     is    an
$\varepsilon$-independent function, is the following for the state
$\Psi^t$,
$$
<f(\varepsilon \sum_m     a^+_ma^-_m)> \to_{\varepsilon\to 0}
f(\varepsilon \sum_m     \varphi^*_m\varphi_m).
$$
This means that the average value of $\varepsilon N$ is
$\sum_{m}\varphi^+_m\varphi^-_m$. To make sure that the $N$-th
component is not exponentially small, one should set
$$
\sum_{m}\varphi^*_m\varphi_m = 1.
\l{b7*}
$$
Let the classical solution to eq.\r{b6} be
$$
\varphi_k^t = \varphi_k e^{-i\Omega t},
\l{b7}
$$
while the $n$-th component of $Y^t$ depend on $t$ as
$$
Y^t_n = Y_n e^{-in\Omega t - i\omega t}.
\l{b8}
$$
It follows from the relation
$$
e^{-i\Omega t\sum_m     a^+_ma^-_m}
a_k^{\pm}
e^{i\Omega t\sum_m     a^+_ma^-_m}
= a_k^{\pm} e^{\mp i\Omega t}
$$
that in this case
$$
U_{\varphi^t,0} =
e^{-i\Omega t\sum_m     a^+_ma^-_m}
U_{\varphi^0,0}
e^{i\Omega t\sum_m     a^+_ma^-_m}.
$$
Eq.\r{b8} means that the vector $e^{i\omega t}
e^{i\Omega t\sum_m a^+_ma^-_m}Y^t$ is time-independent. Therefore, the
$N$-th component of the state $U_{\varphi^t,0}Y^t$ behaves as
$e^{-i(N\Omega+\omega)t}$. Eq.\r{b5*} implies $S=(\Omega -
H(\varphi^*,\varphi))t$, so that the $N$-th component of the vector
\r{b4} depends on $t$ as
$$
\Psi^t_N =\Psi^0_N e^{-i(NH(\varphi^*,\varphi)+\omega)t}
\l{b8*}
$$
and gives us an eigenvector of the $N$-particle Hamiltonian with the
energy
$$
E=NH(\varphi^*,\varphi)+\omega
$$
up to $O(1/N)$.

The next observation is that if one chooses the vector $Y$ as
$$
Y=\sum_m (a_m^+\varphi_m + a_m^-\varphi_m^* +
\sqrt{\varepsilon}\varphi_m^*\varphi_m) X,
$$
the $N$-th component of the expression \r{b4} will vanish. This is a
straightforward corollary of canonical commutation relations \r{b4*}
and the condition \r{b7*}. Since the vector $Y$ is defined up to
$O(\sqrt{\varepsilon})$, one can conclude that if the more weak
condition
$$
Y^t=
e^{-i\Omega t\sum_m a^+_ma^-_m-i\omega t} Y +
 \sum_m (a_m^+\varphi_m + a_m^-\varphi_m^*)X^t
\l{b9}
$$
for some time-dependent vector $X^t$ is satisfied instead of
eq.\r{b8}, the $N$-particle wave function will be also stationary,
so that eq.\r{b8*} will be valid up to $O(\sqrt{\varepsilon})$.

In order to avoid appearence of the Fock space vector $X^t$, it is
convenient to consider the generalized state vector
$$
Z^t= \delta(\sum_m (a_m^+\varphi_m + a_m^-\varphi_m^*))) Y^t.
\l{b10}
$$
The vector $Y^t$ is defined from the relation \r{b10} not uniquely but
up to the state vector of the form
$ \sum_m (a_m^+\varphi_m + a_m^-\varphi_m^*)X^t$. However, different
vectors $Y^t$ obeying eq.\r{b10} determine the same $N$-particle state
$P_NU_{\varphi}Y$ up to $O(\sqrt{\varepsilon})$. This means that the
vector $Z^t$ specifies the asymptotic formula for the $N$-particle
state uniquely.

Eq.\r{b9} implies that the vector $Z$ should obey the following
stationary equation
$$
H_2 Z =\omega Z
\l{b11}
$$
and the constraint equation:
$$
 (\sum_m (a_m^+\varphi_m + a_m^-\varphi_m^*))Z^t=0.
\l{b12}
$$
Thus, one should investigate the Schrodinger equation with the
quadratic Hamiltonian for the constrained system. Analogously to
subsection A.2, introduce new creation and annihilation operators
being linear combinations of the operators $a^{\pm}$:
$$
A_k^+= \sum_l (G^{(k)*}_l a_l^+
 - F^{(k)*}_l a_l^-),
$$
$$
\l{b13}
$$
$$
A_k^-= \sum_l (G^{(k)}_l a_l^-
 - F^{(k)}_l a_l^+).
$$
Such operators obey eq.\r{cr2} if the variation system is satisfied:
$$
-\omega_k F^{(k)}_m =
\sum_l
\left(
\frac{\partial^2H}{\partial \varphi^*_m\partial
\varphi_l} -\Omega \delta_{ml}
\right)  F^{(k)}_l
+ \sum_l\frac{\partial^2H}{\partial \varphi^*_m\partial
\varphi^*_l}G^{(k)}_l,
$$
$$
\l{b14}
$$
$$
\omega_k G^{(k)}_m =
\sum_l
\left(
\frac{\partial^2H}{\partial \varphi_m\partial
\varphi_l^*} -\Omega \delta_{ml}
\right)  G^{(k)}_l
+ \sum_l \frac{\partial^2H}{\partial \varphi_m\partial
\varphi_l}F^{(k)}_l
$$
The ground state vector of eq.\r{b11} is to be found from the
conditions:
$$
A_k^- Z=0.
\l{b15}
$$
Eqs.\r{b15} and \r{b12} determine the state uniquely if the set of
vectors $\varphi^*$, $G^{(1)}$, $G^{(2)}$,... is complete. The new
vacuum vector will be of the Gaussian type
$$
Z=\exp \left(
\frac{1}{2} \sum_{lm} a^+_l R_{lm} a^+_m
\right)|0>,
\l{b16}
$$
where the operator $R$ entering to the quadratic form is defined from
the relations:
$$
R \varphi^* = -\varphi, \qquad RG^{(k)} = F^{(k)}.
$$
The fact the the set $G^{(1)}$, $G^{(2)}$,... may be not complete is
important. This means that the non-stationary analog of eq.\r{b14} may
have one linearily growing at $t$ solution. If we considered the more
strong relation on $Y$, eq.\r{b8}, we would be unable to construct an
asymptotics in such a case.

One of the solutions to eq.\r{b10} on the vector $Y$ is the following:
$$
Y=\exp \left(
\frac{1}{2} \sum_{lm} a^+_l (R_{lm}+\varphi_l\varphi_m) a^+_m
\right)|0>.
\l{b17}
$$
One should then ensure that the expression \r{b17} really determines
the Fock space vector, i.e. that \cite{Ber} $\sum_{mn} |R_{mn}|^2<
\infty$ and $||M||<1$, where the matrix of the operator $M$ is
$M_{mn}=R_{mn}+\varphi_m\varphi_n$. Since $MG^{(k)}=-F^{(k)}$, one
should require that
$$
||G^{(k)}|| > ||F^{(k)}||,
\l{b18}
$$
this rule allows us to select one of two frequences, $\omega$ or
-$\omega$.

Excited states are expressed by eq.\r{b19}, while their energies have
the form \r{b20}. We see that the complex-WKB approach allows us to
find asymptotic spectrum of the $N$-particle Hamiltonian which
corresponds to periodic solutions \r{b7} of eq.\r{b6}.
For more details, see \cite{MS0}
}

\end{document}